\documentclass[sigconf]{acmart}

\newcommand{\name}{ByteDefender}
\AtBeginDocument{%
  }

\copyrightyear{2025}
\acmYear{2025}
\setcopyright{cc}
\setcctype{by}
\acmConference[CCS '25]{Proceedings of the 2025 ACM SIGSAC Conference on Computer and Communications Security}{October 13--17, 2025}{Taipei, Taiwan}
\acmBooktitle{Proceedings of the 2025 ACM SIGSAC Conference on Computer and Communications Security (CCS '25), October 13--17, 2025, Taipei, Taiwan}\acmDOI{10.1145/3719027.3765158}
\acmISBN{979-8-4007-1525-9/2025/10}
\usepackage{booktabs}
\usepackage{longtable}
\usepackage{enumitem} 
\usepackage{subcaption}
\usepackage{multirow}
\usepackage{xcolor}
\usepackage{soul}
\usepackage{listings}
\usepackage{xspace}

\usepackage{tabularx}
\usepackage{ragged2e}
\usepackage{array}

\usepackage{graphicx}

\newcommand{\revedit}[1]{{#1\xspace}}
\definecolor{instructioncolor}{rgb}{0.0, 0.0, 1.0} % Red

\lstdefinestyle{styleByteCode}{
    basicstyle=\ttfamily\footnotesize,
    frame=single, % Frame around the code
    breaklines=false, % Break long lines into multiple lines
    numbers=left,
    numberstyle=\tiny\color{gray},
    moredelim=[is][\color{instructioncolor}]{\%}{\%}
}

\lstdefinestyle{styleJS}{
    basicstyle=\ttfamily\footnotesize,
    keywordstyle=\color{blue},
    stringstyle=\color{red},
    commentstyle=\color{gray},
    morecomment=[l]{//},
    morecomment=[s]{/*}{*/},
    morestring=[b]",
    morestring=[b]',
    breaklines=true,
    showstringspaces=false,
    numbers=left,
    numberstyle=\tiny\color{gray},
    frame=single
}

\begin{document}

% \title{ByteDefender: Leveraging V8 Engine Generated Bytecodes to Identify Fingerprinting Functions Using Transformer Classifier}
% Igor's title proposal
\title{Byte by Byte: Unmasking Browser Fingerprinting at the Function Level Using V8 Bytecode Transformers}

\author{Pouneh Nikkhah Bahrami}
% \orcid{X}
\additionalaffiliation{%
  \institution{Google}
  \city{Sunnyvale}
  \state{California}
  \country{USA}
}
\affiliation{%
  \institution{University of California, Davis}
  \city{Davis}
  \state{California}
  \country{USA}}
\email{pnikkhah@ucdavis.edu}

\author{Dylan Cutler}
\affiliation{%
  \institution{Google}
  \city{Cambridge}
  \state{Massachusetts}
  \country{USA}}
\email{dylancutler@google.com}

\author{Igor Bilogrevic}
\affiliation{%
  \institution{Google}
  \city{Zurich}
  \country{Switzerland}}
 \email{ibilogrevic@google.com}

\renewcommand{\shortauthors}{Pouneh Bahrami, Dylan Cutler, and Igor Bilogrevic}

\begin{abstract}
Browser fingerprinting enables persistent cross-site user tracking via subtle techniques that often evade conventional defenses or cause website breakage when script-level blocking countermeasures are applied.
Addressing these challenges requires detection methods offering both function-level precision to minimize breakage and inherent robustness against code obfuscation and URL manipulation.

We introduce ByteDefender, the first system leveraging V8 engine bytecode to detect fingerprinting operations specifically at the JavaScript function level.
A Transformer-based classifier, trained offline on bytecode sequences, accurately identifies functions exhibiting fingerprinting behavior.
We develop and evaluate lightweight signatures derived from this model to enable low-overhead, on-device matching against function bytecode during compilation but prior to execution, which only adds a 4\% (average) latency to the page load time.
This mechanism facilitates targeted, real-time prevention of fingerprinting function execution, thereby preserving legitimate script functionality.
Operating directly on bytecode ensures inherent resilience against common code obfuscation and URL-based evasion.
Our evaluation on the top 100k websites demonstrates high detection accuracy at both function- and script-level, with substantial improvements over state-of-the-art AST-based methods, particularly in robustness against obfuscation.
ByteDefender offers a practical framework for effective, precise, and robust fingerprinting mitigation.

\end{abstract}

\begin{CCSXML}
<ccs2012>
   <concept>
       <concept_id>10002978.10003029.10011150</concept_id>
       <concept_desc>Security and privacy~Privacy protections</concept_desc>
       <concept_significance>500</concept_significance>
       </concept>
 </ccs2012>
\end{CCSXML}

\ccsdesc[500]{Security and privacy~Privacy protections}

\keywords{Browser Fingerprinting, Bytecode Analysis, Function-level Detection, Transformer Models, Obfuscation Resilience}


\maketitle

\section{Introduction}
Modern JavaScript APIs grant web applications powerful capabilities, from interactive graphics to real-time communication but also expose system details that can be exploited for \emph{browser fingerprinting}.
This technique involves gathering a combination of device and browser characteristics, often through subtle variations in API responses or rendering outputs (e.g., Canvas~\cite{mowery2012pixel}, AudioContext~\cite{das2018web}, fonts~\cite{cao2017cross, vastel2018fp}), to create stable and quasi-unique identifiers. 
While sites may use these techniques for safety-oriented purposes (e.g. anti-fraud, anti-abuse, bot detection) \cite{bursztein2016picasso, durey2021fp, laperdrix2020browser,senol2024double, kalantari2024browser} and for enhancing security by identifying returning devices without traditional authentication, they can also be re-purposed for pervasive, cross-site user tracking~\cite{englehardt2016online, liu2025first, pugliese2020long, vastel2018fp, eckersley2010unique, iqbal2021fingerprinting}.
This tracking capability, often operating statelessly and without clear user awareness or control~\cite{appleAntiFingerpriting, w3cAntiFingerpriting}, raises significant privacy concerns since it can bypass standard privacy-enhancing tools and user preferences~\cite{iqbal2021fingerprinting}.

Mitigating the privacy risks associated with fingerprinting, \emph{without} undermining its safety-oriented uses or breaking website functionality, remains an open challenge. 
Existing countermeasures often fall short, as fine-grained defenses (e.g. adding noise to API outputs~\cite{randomization, datta2018effectiveness}, normalizing values~\cite{cao2017cross, tormitigations,bravemitigations}, or restricting API access~\cite{webkitBlocking, w3mitigations}) could inadvertently disrupt critical fingerprinting applications (e.g., security checks) or break essential website features~\cite{vastel2018fp}. 
Furthermore, scripts may circumvent such defenses~\cite{vastel2018fp} or embrace newly developed fingerprinting vectors~\cite{bahrami2021fp}.

Coarse-grained methods, primarily URL-based filter lists~\cite{disconnect, privacybadger, englehardt2016online}, struggle to differentiate between scripts using fingerprinting for cross-site tracking versus non-tracking purposes. 
They also suffer from delays in updates and are vulnerable against common evasion techniques like CNAME cloaking~\cite{dao2021cname, Dimova2021TheCO}, URL randomization~\cite{wang2016webranz}, and code obfuscation~\cite{moog2021statically, ngan2022nowhere, skolka2019anything}. 
Critically, blocking entire scripts often leads to web breakage, especially problematic when scripts mix essential functions with fingerprinting capabilities~\cite{amjad2024blocking}, and is ineffective against inline scripts \cite{iqbal2021fingerprinting}.

Machine learning-based methods aim to automate the identification of fingerprinting behaviors by leveraging both static and dynamic features of JavaScript code.
Static analysis approaches using ASTs~\cite{rizzo2018machine, van2018detection, iqbal2021fingerprinting} have demonstrated effectiveness but remain susceptible to obfuscation techniques~\cite{ngan2022nowhere, skolka2019anything}. 
Recent work by Ghasemisharif and Polakis~\cite{ghasemisharif2023read} utilized V8 bytecode to detect general \emph{tracking} scripts at the \emph{script-level}, showing resilience against obfuscation.
Amjad et al.~\cite{amjad2024blocking} porposed a \emph{function-level} detection framework for \emph{tracking} behavior using dynamic execution context graphs. 
However, these approaches encounter three key limitations in addressing the fingerprinting dilemma.
First, script-level analysis~\cite{ghasemisharif2023read, iqbal2021fingerprinting} is too coarse to isolate tracking behaviors embedded within multi-purpose scripts, risking disruption of core functionality or failing to block trackers.
Second, dynamic context analysis~\cite{amjad2024blocking}, while finer-grained, can be complex and can introduce runtime overhead, making efficient \emph{pre-execution} intervention difficult. 
Third, neither~\cite{amjad2024blocking, ghasemisharif2023read} specifically isolates fingerprinting behaviors; instead, both conflate stateless (fingerprinting) and stateful (traditional) tracking techniques.
Yet, these two forms of tracking differ fundamentally in their mechanisms and objectives, and a unified detection approach is likely to exhibit mixed effectiveness.

We argue that addressing the privacy risks of fingerprinting requires a solution that is precise (targeting individual functions), robust (resilient to evasion), and proactive (acting \emph{before} execution). 
We introduce \name, the first system using V8 engine bytecode for function-level classification specifically aimed at identifying fingerprinting functions prior to execution. 
By analyzing the bytecode of individual functions during JavaScript compilation, \name\,achieves fine granularity and inherits robustness against source-level obfuscation~\cite{ngan2022nowhere, rozi2020deep}.

{\sloppy Distinct from dynamic context methods~\cite{amjad2024blocking, iqbal2021fingerprinting}, \name\,focuses on the static bytecode patterns within functions, enabling lightweight, pre-execution detection. 
We employ a Transformer-based classifier~\cite{vaswani2017attention, devlin2018bert, kenton2019bert}, trained offline to recognize specific bytecode behaviors associated with fingerprinting techniques~\cite{englehardt2016online, iqbal2021fingerprinting}.
\name\,can identify fingerprinting functions for efficient on-device matching against known fingerprinting signatures, enabling it to preemptively block or alter the execution of fingerprinting functions and mitigate potential privacy leakage without disrupting entire scripts.
This targeted approach minimizes web breakage compared to script-level methods~\cite{ghasemisharif2023read, iqbal2021fingerprinting} and allows core website functionality to proceed without disruption.\par}

\name\,provides a practical pathway to effectively mitigate the privacy risks associated with browser fingerprinting while better preserving web compatibility. Our main contributions are:
\begin{enumerate}
    \item Introduction of the first machine-learning method using V8 engine bytecode for browser fingerprinting detection specifically at the JavaScript function level. Our approach, utilizing a Transformer architecture~\cite{vaswani2017attention} to classify static bytecode sequences, achieves high accuracy (98.9\%), precision (84.0\%), and recall (85.1\%) without reliance on dynamic execution context~\cite{amjad2024blocking} or predefined API lists~\cite{iqbal2021fingerprinting}, while demonstrating robustness against evasion techniques. 
    \item Large-scale evaluation on 100,000 websites, showing high accuracy (99.7\%) and significant improvements over AST-based methods~\cite{iqbal2021fingerprinting, rizzo2018machine, van2018detection} for script-level fingerprinting detection, especially on obfuscated JavaScript code~\cite{moog2021statically, skolka2019anything} and origin/URL manipulation~\cite{dao2021cname, Dimova2021TheCO, wang2016webranz}.
    \item Development of a practical, low-overhead mechanism that only adds 4\% latency (on average) to the page load time for on-device detection and prevention of fingerprinting functions during JavaScript compilation, \emph{before} execution occurs.
    \item Public release of the \name, implementation, and codebase to facilitate reproducibility and further research in bytecode-based fingerprinting mitigation.
\end{enumerate}

The remainder of this paper is structured as follows. 
Section~\ref{sec:background} presents background on JavaScript execution, browser fingerprinting, existing countermeasures, and related work. 
Section~\ref{sec:\name} details the design and implementation of \name, data collection, feature representation, and classifier architecture.
Section~\ref{sec:experimental_setup} describes the experimental setup, dataset characteristics, and implementation of our classifier. 
Section~\ref{sec:evaluation} presents our evaluation results, including accuracy metrics, robustness analysis against obfuscation, comparison with state-of-the-art methods, and the performance overhead of our model in the browser.
Section~\ref{sec:limitations} discusses limitations of our approach.
Section~\ref{sec:conclusion} concludes the paper.

\section{Background and Related Work}\label{sec:background}

This section provides background on the V8 JavaScript execution pipeline, browser fingerprinting techniques and countermeasures, and positions our work with respect to prior research in browser fingerprinting detection.

\subsection{JavaScript Execution and V8 Bytecode}\label{sec:background_JS_pipeline}
\begin{figure}[!thb]
    \centering
    \includegraphics[width=\linewidth]{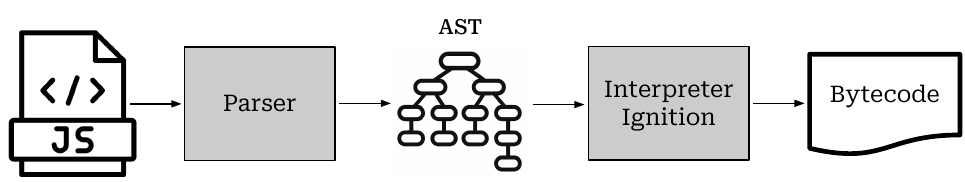} 
    \caption{Simplified V8's compilation pipeline}
    \label{fig:v8_compiler_pipeline}
\end{figure}

JavaScript execution in modern browsers follows a multi-stage pipeline implemented within the V8 engine~\cite{v8Pipeline,v8PipelineFranziska}.
As illustrated in Figure~\ref{fig:v8_compiler_pipeline}, raw JavaScript source code is first parsed into an Abstract Syntax Tree (AST), which represents the code's syntactic structure.
V8's interpreter, Ignition, then compiles this AST into bytecode – a platform-independent, intermediate representation optimized for quick generation and execution.
This bytecode serves as the input for further stages, including interpretation or Just-In-Time (JIT) compilation into optimized machine code by compilers (e.g. TurboFan) for frequently executed functions.
Notably, V8 often employs lazy compilation, generating bytecode for functions only when they are needed~\cite{amjad2024blocking}. 

{\sloppy The bytecode generated by Ignition is compact and not directly human-readable.
However, built-in tools allow disassembly of the bytecode into a sequence of readable instructions such as \texttt{LdaConstant}, \texttt{CallProperty1}, and \texttt{JumpIfFalse}.
\name\,analyzes such a sequence of bytecode instructions produced by Ignition, prior to execution or further optimization.\par}

\subsection{Browser Fingerprinting}

Browser fingerprinting is a technique for gathering device and browser-specific information through various APIs and configurations to create a quasi-unique identifier. 
Unlike traditional stateful tracking via cookies, fingerprinting is typically stateless and less apparent to users.
Common vectors include probing the Canvas API~\cite{mowery2012pixel}, AudioContext~\cite{das2018web}, font availability and rendering~\cite{cao2017cross,vastel2018fp}, WebGL capabilities \cite{wu2019rendered,mowery2012pixel}, WebRTC interfaces~\cite{fifield2016fingerprintability,reiter2017webrtc}, and querying numerous `navigator' properties or system settings. 
A combination of these attributes can derive an identifier with high uniqueness and stability over time.

While acknowledged by browser vendors~\cite{appleAntiFingerpriting,googleAntiFingerpriting,firefoxAntiFingerpriting} and standard bodies~\cite{w3cAntiFingerpriting} as posing privacy risks when used for cross-site tracking~\cite{englehardt2016online}, fingerprinting techniques also have safety-oriented applications in areas such as fraud detection and device authentication~\cite{bursztein2016picasso,durey2021fp, laperdrix2020browser,senol2024double,kalantari2024browser}. 
This dual-use nature makes mitigation complex, as overly broad countermeasures can break such security and anti-abuse mechanisms.

Existing mitigation techniques struggle to effectively address the diverse implementation strategies of fingerprinting. 
Fine-grained API defenses, such as randomization~\cite{randomization, datta2018effectiveness}, normalization~\cite{cao2017cross, tormitigations,bravemitigations}, or blocking~\cite{webkitBlocking}, could break functionality or can be circumvented~\cite{vastel2018fp, datta2018effectiveness, bahrami2021fp}. 
Coarse-grained filter lists~\cite{englehardt2016online, privacybadger, disconnect} are slow to update, are easily evaded by CNAME cloaking~\cite{dao2021cname, Dimova2021TheCO}, URL randomization~\cite{wang2016webranz}, or obfuscation~\cite{skolka2019anything, moog2021statically, ngan2022nowhere}, and cause breakage in mixed scripts~\cite{iqbal2021fingerprinting, amjad2024blocking}. 
Machine learning offers automation but faces trade-offs.

Static analysis using Abstract Syntax Trees (ASTs), explored in~\cite{rizzo2018machine, van2018detection} and adopted by FP-Inspector~\cite{iqbal2021fingerprinting}, analyzes the structural representation of JavaScript code. 
These methods extract features such as specific API calls (e.g., \texttt{toDataURL}, \texttt{measureText}), structural patterns (e.g., sensitive APIs invoked within loops) and various complexity metrics.
Although ASTs offer a structured view of code semantics, they are highly sensitive to syntactic changes, making them vulnerable to even minor obfuscation.
Code obfuscation techniques~\cite{skolka2019anything, moog2021statically, ngan2022nowhere}, including renaming variables/functions, inserting dead code, or flattening control flow, drastically alter the AST structure.
Moreover, ASTs require extensive post-processing to extract usable features and are not directly suitable as input to learning models.
These limitations make AST-based techniques ill-suited for real-time, proactive fingerprinting detection and hinder their ability to generalize to obfuscated or novel scripts.

Dynamic analysis addresses some of these issues by observing runtime behavior.
FP-Inspector~\cite{iqbal2021fingerprinting} supplemented its static analysis with dynamic features obtained by executing scripts. 
More recently, NoT.js~\cite{amjad2024blocking} employed extensive dynamic analysis using browser instrumentation (via the DevTools Protocol) to build detailed execution graphs.
These graphs capture the sequence of function calls (call stack) and the data accessible at each point (scope chain) when network requests or other specific events occur. 
NoT.js uses features derived from this graph (e.g., node types, connectivity) to train a classifier identifying tracking functions, focusing on their role in initiating tracking requests, and subsequently generates surrogate scripts. 
While powerful for understanding runtime execution flow and achieving function-level granularity for tracking, this approach requires significant browser instrumentation, potentially incurring runtime overhead. 
Crucially, detection happens during or after execution, making preemptive blocking difficult. 
Furthermore, its features are derived from the effects of execution (e.g., API calls made, network requests sent), whereas \name\,analyzes the intrinsic structure (bytecode) of the function itself before execution to predict its fingerprinting potential based on learned patterns corresponding to fingerprinting API usage.

Script-level bytecode analysis, introduced by Ghasemisharif and Polakis~\cite{ghasemisharif2023read} inspired by~\cite{rozi2020deep} for general tracking/ad detection, established V8 bytecode as a robust feature source resilient to obfuscation.
They treated entire JavaScript files as documents and applied text classification techniques (e.g., DPCNN and FastText) to the concatenated bytecode sequences of all functions within a script. 
While demonstrating bytecode's advantages over ASTs, this script-level granularity introduces two major limitations.
First, it conflates fundamentally distinct forms of tracking: stateless and stateful, resulting in mixed detection effectiveness.
Second, labeling the entire script as tracking can lead to coarse blocking, mirroring the breakage issues seen with filter lists~\cite{iqbal2021fingerprinting} when legitimate and tracking functions coexist in a single script.

\name\, integrates the strengths of these related areas while overcoming their key limitations for the specific task of fingerprinting detection. 
It adopts V8 bytecode as the core feature, inheriting the demonstrated obfuscation resilience~\cite{ghasemisharif2023read, rozi2020deep} that AST-based methods lack~\cite{iqbal2021fingerprinting, rizzo2018machine, van2018detection}.
Crucially, unlike script-level bytecode analysis~\cite{ghasemisharif2023read, iqbal2021fingerprinting}, \name\,  operates at the function-level. 
By analyzing the bytecode sequence of each individual function and complementing it with execution traces of the function, the approach constructs a novel function-level ground truth that provides the fine granularity needed to identify specific fingerprinting behaviors within mixed-purpose scripts, thereby minimizing the risk of web breakage associated with blocking entire files. 
Distinct from dynamic analysis approaches such as NoT.js~\cite{amjad2024blocking}, \name\, performs static analysis on the function's bytecode during compilation and before execution. 
This allows for lightweight, proactive intervention, preventing potential information leakage before the fingerprinting function runs, a significant advantage over methods requiring runtime observation.
Instead of relying on dynamic call graphs or execution context, \name\, trains a Transformer model~\cite{vaswani2017attention} to directly recognize the intrinsic bytecode patterns and instruction sequences to identify signatures for compilation-time blocking.
This focus on the code's static representation at the bytecode level, combined with function-level granularity and pre-execution timing, provides a unique approach specifically tailored to proactively mitigating fingerprinting risks.

\section{\name: Function-Level Fingerprinting Detection}\label{sec:\name}

\begin{figure*}[!thb]
    \centering
    \includegraphics[width=0.9\linewidth]{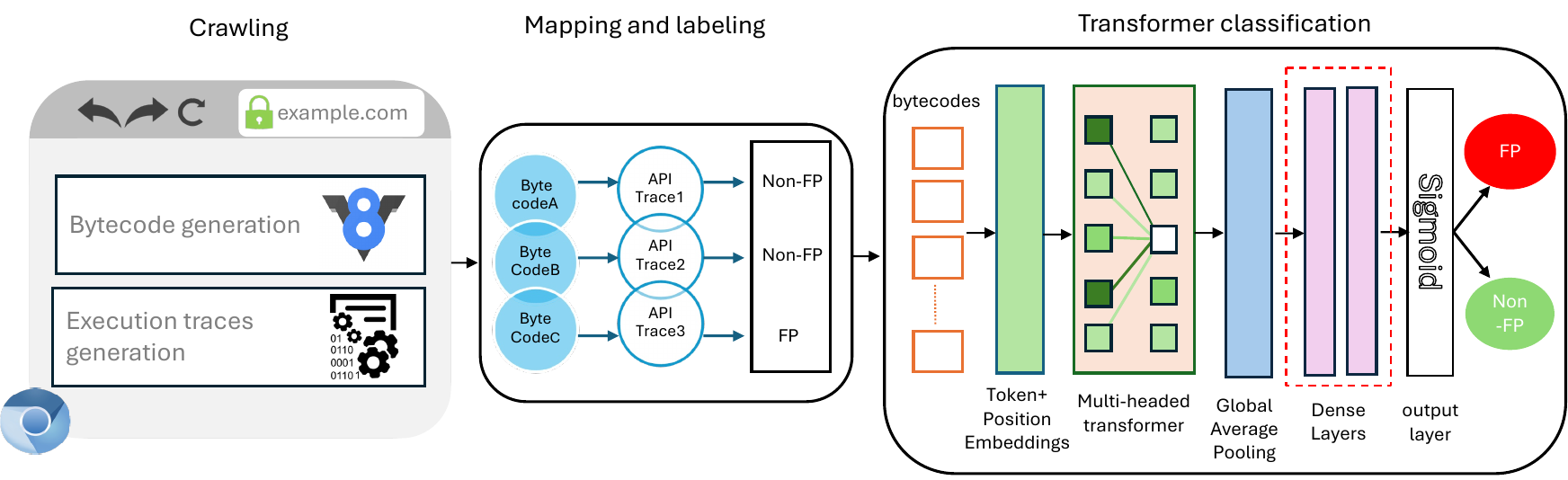}
    \caption{\name: (1) Web crawling is performed using a Chromium browser with an instrumented V8 engine to extract function-level bytecodes and execution traces via a custom extension. (2) Execution traces of functions are mapped to their corresponding function bytecodes and labeled as Fingerprinting (FP) or Non-Fingerprinting (Non-FP). (3) A transformer-based classifier is trained on labeled bytecodes.}
    \label{fig:ByteDefender_arcitechture}
    \vspace{-0.1 in}
\end{figure*}

In this section we detail the design and implementation of \name{}. 
The core methodology involves collecting function-level bytecode and corresponding execution traces, using these traces to establish ground truth labels for fingerprinting activity, mapping these labels to the corresponding bytecodes, and training a Transformer-based classifier on the labeled bytecode sequences. 
Figure~\ref{fig:ByteDefender_arcitechture} provides a high-level overview of \name{}.

A primary challenge in this work is the lack of pre-existing ground truth labels for function bytecode as it relates to fingerprinting. 
To overcome this, we leverage execution trace analysis to infer labeling at the function level.
Prior research confirms that fingerprinting techniques rely on specific patterns of Web API usage~\cite{englehardt2016online, das2018web, iqbal2021fingerprinting}. 
For instance, canvas fingerprinting typically involves calls to \texttt{fillText()} followed by \texttt{toDataURL()} or \texttt{getImageData()}~\cite{mowery2012pixel}. 
{\sloppy Similarly, AudioContext fingerprinting uses a characteristic sequence including methods such as \texttt{createOscillator} and \texttt{startRendering}~\cite{das2018web}.\par}
By identifying known API interaction patterns in execution traces captured during web crawls, we can heuristically label functions that exhibit fingerprinting behavior.
This labeled trace data then serves as the basis for training our bytecode classifier.

Our approach comprises several key steps: (1) instrumenting the V8 engine to extract detailed function-level bytecode along with necessary metadata (script URL, script Id, function name); (2) using a browser extension to capture fine-grained execution traces, focusing on high-entropy API calls; (3) collecting bytecode and trace data via automated web crawling; (4) applying heuristics based on known fingerprinting techniques to label functions in the trace data; (5) mapping these labels to their corresponding function bytecode sequences; and (6) training and evaluating classifiers on the resulting labeled bytecode dataset.

\subsection{Instrumented V8 Bytecode Collection}\label{sec:v8}

As we discussed in Section~\ref{sec:background_JS_pipeline}, V8 converts JavaScript source code into bytecode via the Ignition interpreter. 
While standard V8 tools (such as \texttt{d8} and Node.js with the \texttt{--print-bytecode} flag) expose bytecode, they do not provide essential metadata that links the bytecode to its originating script URL, script ID, or function name.
Listing \ref{lst:original_bytecode} displays the original bytecode for the JavaScript code found in Listing \ref{lst:raw_script}, which was generated by V8 during the compilation process. 
As shown in Listing \ref{lst:original_bytecode}, there is a lack of detailed information about the script and function being compiled. 
This metadata is crucial for mapping ground-truth labels derived from execution traces (which contain script/function context) to the corresponding bytecode sequences.
\begin{lstlisting}[style=styleJS, caption={A Script source code which performs browser fingerprinting}, label={lst:raw_script}]
function gatherFingerprint() {
    var fingerprint = {
        userAgent: navigator.userAgent,
        language: navigator.language,
        platform: navigator.platform,
        screenResolution: `${screen.width}x${screen.height}`
    };
    return fingerprint;
}
gatherFingerprint();
\end{lstlisting}
\begin{lstlisting}[style=styleByteCode, caption={A truncated and original bytecode generated by V8 for the script shown in Listing~\ref{lst:raw_script}}, label={lst:original_bytecode}]
[generated bytecode (0x16ce00098fb5 <SharedFunctionInfo>)]
Bytecode length: 79
Parameter count 1
Register count 4
Frame size 32
0x7ff2001400c8 @ 0 :  85 00  %CreateObjectLiteral [0], [0]%
0x7ff2001400cc @ 4 :  cd     %Star1%
0x7ff2001400cd @ 5 :  23 01  %LdaGlobal [1], [1]%
0x7ff2001400d0 @ 8 :  cc     %Star2%
0x7ff2001400d1 @ 9 :  33 f7  %GetNamedProperty r2, [2]%
0x7ff2001400d5 @ 13 : 3a f8 %DefineNamedOwnProperty r1%
0x7ff2001400d9 @ 17 : 23 01 %LdaGlobal [1], [1]%
0x7ff2001400dd @ 21 : 33 f7 %GetNamedProperty r2, [3], [7]%
...
\end{lstlisting}

To overcome this limitation, we instrument the V8 source code. 
Specifically, we modify the \texttt{DoFinalizeJobImpl()} function within \texttt{interpreter.cc} to access and output the \texttt{scriptId}, \texttt{scriptURL}, and \texttt{functionName} associated with the \texttt{SharedFunctionInfo} object, alongside the generated bytecode instructions by \texttt{BytecodeArray} class.
We call \texttt{BytecodeArray::Disassemble()} to convert the binary format of raw bytecodes to the corresponding mnemonic name that represent each bytecode instruction.

{\sloppy Furthermore, to manage the verbosity of bytecode output for large scripts, we modify V8's bytecode printing functions (\texttt{NameForRuntimeId}, \texttt{Decode} in \texttt{bytecode-decoder.cc}, and \texttt{Disassemble} in \texttt{bytecode-array.cc}) to log only the sequence of bytecode instruction names, omitting memory addresses, operands, comments, and bytecode offsets.\par}
While this choice simplifies feature extraction and reduces noise, it also introduces a potential limitation: the loss of fine-grained semantic context, including constants, property names, or literal values (e.g., "userAgent", "canvas", or "toDataURL"), which may help in distinguishing between non-fingerprinting and fingerprinting behavior.
However, this design decision is intentional and aligned with our goal of achieving generalizable, obfuscation-resilient fingerprinting detection. 
Many common obfuscation techniques (e.g., string encoding, variable renaming, and control flow flattening) target operands and syntactic constructs, but tend to preserve the underlying operational structure captured in the bytecode instruction sequences. 
By focusing solely on opcode patterns and abstracting away operands, our model learns behavioral representations that are more invariant to obfuscation.
Listing~\ref{lst:intrumented_bytecode} shows the output format from our instrumented V8 for the script in Listing~\ref{lst:raw_script}, now including the necessary metadata.

To handle concurrent compilation tasks within V8, which uses multiple processes and threads, we ensure unique output filenames for each compilation unit by incorporating thread ID, process ID, and a timestamp into the filename, preventing concurrent write conflicts or file overwrites.

\begin{lstlisting}[style=styleByteCode,caption={Bytecode generated by our instrumented V8 for the script in Listing~\ref{lst:raw_script}}, label={lst:intrumented_bytecode}]
Script URL: \url{https://example.com/fpjs.js}
Script ID: 3
Function name: gatherFingerprint
Bytecode: 
Parameter count 1
Register count 4
Frame size 32
%[CreateObjectLiteral,Star1,LdaGlobal,Star,GetNamedProprty,
DefineNamedOwnProperty,LdaGlobal,Star,GetNamedProperty,
DefineNamedOwnProperty,LdaGlobal,Star2,GetNamedProperty, 
DefineNamedOwnProperty,LdaGlobal,Star3,GetNamedProperty, 
ToString,Star2,LdaConstant,Add,Star2,LdaGlobal,Star3,
GetNamedProperty,ToString,Add,DefineNamedOwnProperty,Mov,
Ldar,Return]%
\end{lstlisting}

\subsection{Execution Trace Collection via Browser Extension}\label{sec:extension}
Complementary to bytecode collection, we capture dynamic execution traces using a custom Chrome browser extension. 
This extension employs the \texttt{Tracing} domain of Chrome DevTools Protocol (CDP) \cite{tracing_CDP} which gathers fine-grained events that provide a detailed timeline of browser activities such as function calls during JavaScript execution.
This data is captured as JSON objects.

The extension operates by attaching to the active browser tab and initiating a CDP tracing session. 
Crucially, for the purpose of generating ground truth labels via established heuristics (Section~\ref{sec:Detecting_fingerprinting_functions}), this tracing session is specifically configured to capture events associated only with "High Entropy APIs" as designated by Chromium~\cite{high_entropi_apis}.
These APIs (Appendix~\ref{sec:appendix1}, Table~\ref{tab:highEntropyAPIs}) expose potentially identifying information and are natively instrumented within Chrome, which makes them detectable via the \texttt{Tracing} domain. 
Consequently, traces are collected only for functions that invoke at least one of these specific APIs.
This focus on high-entropy API calls allows us to effectively apply heuristics from prior research~\cite{englehardt2016online, iqbal2021fingerprinting, das2018web} for labeling purposes.

{\sloppy During page execution, the extension listens for \texttt{Tracing.dataCollected} events corresponding to these high-entropy API calls. 
It parses these events to extract relevant details for each call, including the API identifier, arguments, the URL and function name of the calling script, line/column numbers, and the top-level page URL. 
Traces are filtered to exclude noise from invalid sources (e.g., internal browser pages starting with \texttt{chrome:}, extension URLs, \texttt{file:} URLs) or DevTools activity itself).

The filtered, structured trace data is then serialized to JSON and transmitted to our local server for further analysis and labeling. 
When scripts are parsed, the extension uses the CDP \texttt{Debugger} domain to retrieve and send the raw script source code, along with the script's unique identifier and URL, to the server for processing.\par}

\subsection{Data Collection Framework}\label{sec:data_collection}
We integrate the instrumented V8 browser (Section~\ref{sec:v8}) and the tracing extension (Section~\ref{sec:extension}) into an automated crawling framework built using Selenium WebDriver~\cite{selenium}. 
The framework systematically navigates to the homepages of target websites (Section~\ref{sec:experimental_setup}). 
For each visited page, the instrumented V8 logs function-level bytecode with metadata, while the extension concurrently captures and filters execution traces and script sources. 
To allow sufficient time for dynamic scripts to execute, the crawler interacts with each page for a fixed duration (15 seconds) before proceeding. 
The crawler finally transmits the bytecode logs and serialized trace data to our local analysis servers.

\subsection{Ground Truth Generation via Heuristic Labeling}\label{sec:Detecting_fingerprinting_functions}

As ground truth labels for fingerprinting behavior in bytecode are not available, we generate them heuristically by analyzing the collected execution traces, as described in Section~\ref{sec:extension}.
We leverage the fact that specific fingerprinting techniques are characterized by distinct patterns of high-entropy API calls~\cite{englehardt2016online, iqbal2021fingerprinting}.
Based on established research, we define heuristics for four common fingerprinting categories: Canvas, Canvas Font, Audio Context, and WebRTC. 
Applying these heuristics to the captured traces allows us to label function execution instances as fingerprinting or non-fingerprinting and provides the ground truth for training our bytecode classifiers.
Below we list fingerprinting techniques along with detection heuristics we used to build our initial ground truth of fingerprinting and non-fingerprinting functions.

\textbf{Canvas Fingerprinting Heuristic.}
Inspired by~\cite{mowery2012pixel, iqbal2021fingerprinting}, this technique exploits differences in graphics rendering.
To detect functions employing this technique, all of the following criteria must be met in the execution trace:
\begin{sloppy}
\begin{itemize}[noitemsep,topsep=0pt]
    \item Call to \texttt{fillText()} of \texttt{CanvasRenderingContext2D} or \texttt{OffscreenCanvasRenderingContext2D}.
    
    \item Call to the \texttt{toDataURL} method within the same function to export the Canvas content.
    
    \item The length of text written via \texttt{fillText()} is at least 10 characters.
    
    \item The function trace avoids invoking the \texttt{save}, \texttt{restore}, or \texttt{addEventListener} methods of the rendering context.
    
\end{itemize}
\end{sloppy}

\textbf{Canvas Font Fingerprinting Heuristic.}
This approach infers installed fonts by measuring text rendering dimensions~\cite{iqbal2021fingerprinting}, rather than direct browser’s font list enumeration.
To detect functions employing this technique, all of the following criteria must be met:
\begin{sloppy}
\begin{itemize}[noitemsep,topsep=0pt]
    \item Calls to \texttt{CanvasRenderingContext2D.measureText} or \texttt{OffscreenCanvasRenderingContext2D.measureText} occur at least 20 times within the function trace.
    
    \item Calls setting the font property (e.g., \texttt{CanvasRenderingContext2D.font.set}) occur for more than 20 different font values within the function trace.
    
\end{itemize}
\end{sloppy}

\textbf{Audio Context Fingerprinting Heuristic.}
This technique leverages subtle variations in audio signal processing due to hardware or browser differences~\cite{das2018web}.
To detect functions employing this technique, all of the following criteria must be met:
\begin{sloppy}
\begin{itemize}[noitemsep,topsep=0pt]
    \item Calls to characteristic AudioContext API methods for audio creation and manipulation are present (e.g., \texttt{BaseAudioContext.createOscillator}, \texttt{BaseAudioContext.createDynamicsCompressor}, \texttt{OfflineAudioContext.startRendering}, \texttt{AudioNode.connect}).
    
    \item A subsequent call to \texttt{AudioBuffer.getChannelData} occurs within the function trace to extract the processed audio data.
\end{itemize}
\end{sloppy}

\textbf{WebRTC Fingerprinting Heuristic.}
Exploits the exposure of low-level network details via \texttt{RTCPeerConnection}~\cite{fifield2016fingerprintability, reiter2017webrtc, iqbal2021fingerprinting}.
To detect functions employing this technique, all of the following criteria must be met:
\begin{sloppy}
\begin{itemize}[noitemsep,topsep=0pt]
    \item Call to \texttt{RTCPeerConnection.createDataChannel} or \texttt{RTCPeerConnection.createOffer} to initiate SDP generation.
    
    \item Call to \texttt{RTCPeerConnection.setLocalDescription} to access SDP details, potentially containing local IP addresses.
\end{itemize}
\end{sloppy}

\subsection{Mapping Labels to Bytecode Sequences}\label{sec:assiging_lables}
The crucial step is mapping the fingerprinting/non-fingerprinting labels derived from execution traces (Section~\ref{sec:Detecting_fingerprinting_functions}) to the corresponding bytecode sequences collected by the instrumented V8 (Section~\ref{sec:v8}). 
We do so by joining the trace data and the bytecode log data using the unique combination of \textit{script URL}, \textit{script ID}, and \textit{function name} present in both datasets thanks to our instrumentation.
To ensure a consistent dataset, we discard records associated with anonymous functions, as they lack a reliable name for joining. 
If a unique combination of \textit{script URL}, \textit{script ID}, and \textit{function name} for a function's bytecode does not have a corresponding record in the execution traces, the function is labeled as non-fingerprinting. 
Since traces were only captured for functions invoking high-entropy APIs, the absence of a trace implies the function made no such calls, and therefore failed to trigger any of our fingerprinting detection heuristics.
We also discard records with invalid or non-web URLs (e.g., internal V8 metrics, browser extension paths starting with \texttt{chrome-extensions::}, \texttt{v8/}, \texttt{chrome:}, \texttt{file:}) which are not valid URLs. 
This mapping process yields our final dataset: Sequences of function-level bytecode labeled as either fingerprinting or non-fingerprinting. 
We use this labeled dataset to train our classifiers.

\subsection{Bytecode Representation via Word Embeddings}\label{sec:word_embedding}
To leverage machine learning for classifying function behavior based on bytecode, we must first transform the discrete sequences of bytecode instructions (obtained as described in Section~\ref{sec:v8}) into numerical vector representations suitable for model input. 
This process, known as embedding, maps each unique bytecode instruction (our "word" or "token") into a continuous vector space, ideally positioning instructions with similar semantic roles closer together. 
The choice of embedding technique impacts how contextual information and instruction relationships are captured. 
We explore established methods, adapting them for the unique characteristics of V8 bytecode.

\textbf{Static Embeddings (Word2Vec and FastText).} These methods generate a single, fixed vector representation for each unique bytecode instruction, learned from a large corpus of bytecode sequences in an unsupervised way.
\begin{itemize}[noitemsep, topsep=0pt, leftmargin=*]
    \item \textbf{Word2Vec}~\cite{mikolov2013efficient}: We employ the Skip-gram architecture, which learns embeddings by training a shallow neural network to predict the context bytecode instruction (neighboring bytecodes within a defined window) given a target bytecode instruction. 
    For example, given the bytecode sequence `..., LdaGlobal, GetNamedProperty, Star, ...', the model learns vectors such that the embedding for `GetNamedProperty' helps predict `LdaGlobal' and `Star' if they fall within the context window. 
    The weights from the resulting hidden layer are used to form the embedding matrix.
    Key hyperparameters include the embedding dimension (vector size, e.g. 100) and the context window size (e.g. 3 or 5 bytecode instructions before and after).
    While computationally efficient, Word2Vec primarily captures local co-occurrence patterns and assigns the same vector regardless of the instruction's context in a specific sequence~\cite{goldberg2014word2vec}.
    
    \item \textbf{FastText}~\cite{joulin2016bag}: FastText enhances Word2Vec by representing each bytecode instruction not just as a whole unit but also as a bag of character n-grams. 
    For a hypothetical bytecode instruction `LdaConstant', FastText might consider n-grams like `<Ld', `Lda', `daC', `aCo', ..., `ant>', along with the full instruction `<LdaConstant>'. 
    This allows the model to share representations between instructions with similar character patterns (e.g., `LdaConstant' and `LdaGlobal') and potentially generate better embeddings for rare instructions not frequently seen during training. 
    Like Word2Vec, FastText produces static embeddings, but it requires more memory due to storing n-gram vectors.
\end{itemize}

These static embeddings serve as input for our baseline Random Forest classifier (Section~\ref{sec:classifier}). 
We determine optimal hyperparameters (vector dimension, window size, training epochs) through experimentation (details in Section~\ref{sec:experimental_setup}).

\textbf{Contextual Embeddings (Learned by Transformer).} Our primary approach utilizes the Transformer architecture itself to learn dynamic, context-aware embeddings. 
Unlike static methods, the embedding representation for a bytecode instruction changes based on its position and the surrounding bytecodes within a specific function's bytecode sequence.
\begin{itemize}[noitemsep, topsep=0pt, leftmargin=*]
    \item \textbf{Input Representation:} We treat each unique V8 bytecode instruction (e.g., `CreateObjectLiteral', `LdaGlobal', `GetNamedProperty', `Return') as a token in our vocabulary. 
    Since pre-trained Transformer models are trained on natural language (e.g. BERT~\cite{kenton2019bert, devlin2018bert}, GPT~\cite{radford2018improving}), their vocabularies and learned representations are unsuitable for bytecode instructions. 
    Therefore, we initialize a token embedding matrix ($\mathbf{E}_{token} \in \mathbb{R}^{V \times d}$) randomly, where $V$ is the vocabulary size (number of unique bytecode instructions) and $d$ is the embedding dimension.
    \item \textbf{Embedding Calculation:} For a bytecode instruction $i$ at position $j$ in a sequence, its input vector is calculated as $\mathbf{x}_j = \mathbf{E}_{token}[i]$.
    We supplement these embeddings with sinusoidal position encoding~\cite{vaswani2017attention}.
    \item \textbf{Contextualization via Self-Attention:} This sequence of input vectors ($\mathbf{x}_1, ..., \mathbf{x}_L$) is fed into a Transformer block~\cite{vaswani2017attention}. 
    The multi-head self-attention mechanism within the block allows each bytecode instruction's representation to be updated based on its relationship with all other bytecode instructions in the sequence.
    For example, the final representation for a `GetNamedProperty' bytecode instruction will differ depending on whether it follows an `LdaGlobal' (potentially accessing a global object property) or an `LdaSmi' (an unlikely sequence, perhaps indicating unusual code). 
    These learned embeddings are optimized end-to-end specifically for the fingerprinting classification task.
\end{itemize}

This approach allows the model to capture complex dependencies and nuances specific to bytecode sequences used in fingerprinting, offering a richer representation than static embeddings.

\subsection{Fingerprinting Classification Models}\label{sec:classifier}
Using the labeled bytecode sequences and their vector representations, we train classifiers to predict whether a given function exhibits fingerprinting behavior. We evaluate two distinct model architectures:

\textbf{Random Forest (Baseline).} We utilize Random Forest (RF) as a strong traditional machine learning baseline. 
RF is an ensemble method constructing multiple decision trees during training and making a final classification based on which class is predicted by the majority of the trees~\cite{pedregosa2011scikit}.
\begin{itemize}[noitemsep, topsep=0pt, leftmargin=*]
    \item \textbf{Input Features:} RF requires a fixed-size feature vector for each sample (function). 
    We generate this by taking the static embeddings (Word2Vec or FastText, as described in Section~\ref{sec:word_embedding}) learned for each bytecode instruction and computing the element-wise average of the embedding vectors for all tokens present in a given function's bytecode sequence. 
    This results in a single $d$-dimensional vector per function, where $d$ is the embedding dimension.
    \item \textbf{Training:} The RF model is trained on these averaged vectors and their corresponding fingerprinting/non-fingerprinting labels. 
    Key hyperparameters, such as the number of trees, maximum depth of each tree, criteria for splitting nodes (e.g., `gini' or `entropy'), and the number of features considered at each split, are tuned using techniques (e.g., GridSearchCV~\cite{pedregosa2011scikit}) on a validation set. 
    RF models offer relatively fast training and some level of interpretability through feature importance analysis, but the averaging process inherently loses sequential information present in the bytecode.
\end{itemize}

\textbf{Transformer Classifier.} Our primary model is a custom Transformer classifier, leveraging the architecture's proven success in sequence modeling~\cite{vaswani2017attention, devlin2018bert}. 
This deep learning approach processes the entire bytecode sequence directly, preserving order and context.
\begin{itemize}[noitemsep, topsep=0pt, leftmargin=*]
    \item \textbf{Architecture:} As illustrated in Figure~\ref{fig:ByteDefender_arcitechture} (right panel) and detailed below, our model follows a standard Transformer encoder structure adapted for classification:
        \begin{enumerate}[label=\alph*), noitemsep, topsep=0pt]
            \item \textit{Input Embeddings:} Summed learned token embeddings along with positional encoding for the bytecode sequence (Section~\ref{sec:word_embedding}).
            \item \textit{Transformer Encoder Layer:} A multi-head self-attention sub-layer followed by a position-wise fully connected feed-forward network sub-layer.
            Residual connections and layer normalization are applied around each sub-layer~\cite{vaswani2017attention}.
            This allows the model to progressively build richer contextual representations of the sequence.
            \item \textit{Pooling:} A Global Average Pooling layer takes the output sequence from the final Transformer layer ($H \in \mathbb{R}^{L \times d}$) and computes the mean across the sequence length dimension, resulting in a single fixed-size vector ($h \in \mathbb{R}^{d}$) that summarizes the entire function's bytecode context.
            \item \textit{Classification Head:} The pooled vector $h$ is passed through one or more dense (fully connected) layers with ReLU activation functions and dropout~\cite{srivastava2014dropout} for regularization, before a final output layer consisting of a single neuron with a sigmoid activation function.
        \end{enumerate}
    % \item \textbf{Training:} The model is implemented using JAX/TensorFlow~\cite{frostig2018compiling, abadi2015tensorflow} and trained end-to-end using binary cross-entropy loss with the Adam optimizer~\cite{kingma2014adam}. The gradients flow back through the entire network, including the randomly initialized embedding layers, allowing the model to learn representations specifically optimized for distinguishing fingerprinting bytecode patterns. Hyperparameters (number of layers, heads, embedding dimension, feed-forward dimension, learning rate, dropout rate) are tuned on a validation set. While requiring more computational resources for training compared to RF, the Transformer can capture complex long-range dependencies and contextual nuances within the bytecode sequence that are lost in the averaging process used for the RF input.
\end{itemize}
This end-to-end architecture (Figure~\ref{fig:ByteDefender_arcitechture}, right panel) is trained specifically to optimize fingerprinting classification accuracy based on bytecode patterns. Performance comparisons are detailed in Section~\ref{sec:evaluation}.

\section{Experimental Setup and Dataset}\label{sec:experimental_setup}

This section details the experimental environment, the large-scale data collection process used to gather function-level bytecode and execution traces, the data cleaning and labeling pipeline, and the resulting dataset characteristics. 
We also describe the model architectures and hyperparameter tuning strategies employed for our classifiers.

\subsection{Hardware and Crawling Infrastructure}
All data collection was done on an Ubuntu 22.04 server equipped with an Intel Xeon Silver 4216 CPU and 376GB of RAM.
Model training and testing was done with eight TPU v4 units, each with 16 GB of memory.
Data collection utilized an automated web crawler built with Selenium WebDriver, controlling an instrumented version of the Chromium browser (version 123.0.6284.0, Section~\ref{sec:v8}) augmented with our custom tracing extension (Section~\ref{sec:extension}). 

Crawling was conducted from a US-based university network IP address in June 2024.
The crawler visited the homepages of the top 100,000 websites listed in the Tranco top-sites list~\cite{pochat2018tranco}, interacting with each page for 15 seconds to capture dynamic behavior before sending collected data (bytecode logs and serialized traces) to local servers.

\subsection{Dataset Construction and Cleaning}\label{subsec:dataset_construction}
The instrumented V8 engine generates bytecode sequences at the function-level, reflecting the compilation order and lazy execution semantics. 
Concurrently, our extension captures API execution traces.

\textbf{Initial Data Collection.} The raw data collection yielded bytecode and traces corresponding to approximately 21.1 million script instances across the 100k websites, comprising roughly 658 million function instances.
This initial dataset contained significant noise, including functions from scripts with invalid URLs, anonymous functions lacking names, and functions executed using \texttt{eval}. Analysis of the raw execution traces showed that about 9.6M function instances (1.5\% of total) invoked at least one high-entropy API potentially related to fingerprinting.
Of these, 6.6M were anonymous functions (including 2,905 eventually labeled fingerprinting) and 3.1M were named functions (including 4,690 eventually labeled fingerprinting).

\textbf{Data Cleaning and Label Mapping.} To construct a clean dataset suitable for training and evaluation and to map execution trace labels to bytecode sequences, we performed several filtering steps using the methodology in Section~\ref{sec:assiging_lables}:
\begin{enumerate}[noitemsep, topsep=0pt]
    \item \textit{URL Filtering:} Records associated with scripts having invalid or empty URLs were removed.
    This reduced the dataset to ~9.4M scripts and ~656M functions (7,595 fingerprinting functions).
    \item \textit{Anonymous \revedit{and Eval-Loaded} Function Filtering:} 
    \revedit{As described in Section~\ref{sec:assiging_lables}, our labeling process requires mapping dynamic execution traces to their corresponding static bytecode sequences using a key composed of script URL, script ID, and function name. Anonymous and eval-loaded functions lack a stable name, making reliable trace-to-bytecode mapping infeasible. We therefore exclude them from training dataset.}
    % As function names are required for accurate label-to-bytecode mapping, records corresponding to anonymous \revedit{and eval-loaded} functions were excluded.
    %
    This further reduced the dataset to ~4.5M scripts (4,666 fingerprinting) and ~407M functions (4,690 fingerprinting).
    Notably, this step removed a significant portion ($\sim$38\%) of both potential fingerprinting and non-fingerprinting functions.
    \revedit{While this exclusion limits the diversity of our training set, it does not impact deployment: ByteDefender relies solely on bytecode structure and does not require function names. As a result, it can still detect fingerprinting in anonymous or eval-loaded functions if their bytecode exhibits patterns learned from named functions during training.}
    \item \textit{Duplicate Removal:} Duplicate function records (i.e., bytecode sequences) were removed to prevent data contamination between training and testing sets.
    That is, for the function-level evaluation (Section~\ref{subsec:fn_level_eval}) we removed all duplicate function bytecode sequences, whereas for the script-level evaluation (Section~\ref{subsec:script_level_eval}) we removed all duplicate script bytecode sequences.
    This resulted in the final function count.
\end{enumerate}
{\sloppy The final processed dataset statistics are presented in Table~\ref{tab:original_dataset_stats}.
Each record contains the website domain, script ID, script URL, function name, the generated bytecode sequence, and its assigned fingerprinting/non-fingerprinting label derived from trace heuristics (Section~\ref{sec:Detecting_fingerprinting_functions}).\par}

\begin{table}[htbp]
\centering
\caption{Distribution of non-fingerprinting (non-FP) and fingerprinting(FP) scripts and functions across the dataset after successive cleaning steps.}
\label{tab:original_dataset_stats}
\resizebox{\columnwidth}{!}{
\begin{tabular}{l|rr|rr}
\toprule
\textbf{Cleaning Step} & \multicolumn{2}{c|}{\textbf{\# Scripts}} & \multicolumn{2}{c}{\textbf{\# Functions}} \\
\cmidrule(lr){2-3} \cmidrule(lr){4-5}
 & \multicolumn{1}{c}{\textbf{FP}} & \multicolumn{1}{c|}{\textbf{non-FP}} & \multicolumn{1}{c}{\textbf{FP}} & \multicolumn{1}{c}{\textbf{non-FP}} \\
\midrule
Removing invalid/empty scripts & 7,539 & 9,363,941 & 7,595 & 655,930,885 \\
Removing anonymous functions   & 4,666 & 4,456,889 & 4,690 & 407,310,091 \\
Removing repeated rows         & 4,666 & 4,456,889 & 4,670 & 404,877,008 \\
\bottomrule
\end{tabular}
}
\end{table}

% \textbf{Dataset Imbalance:} The final dataset exhibits significant class imbalance, with non-fingerprinting functions vastly outnumbering fingerprinting functions (4,670 fingerprinting vs. ~405M non-fingerprinting).
% %
% This is expected, as fingerprinting behavior, especially involving high-entropy APIs, is relatively rare compared to general script functionality. 
% %
% To mitigate bias towards the majority class during training, we employ oversampling of the positive (fingerprinting) class and undersampling of the negative (non-fingerprinting) class. 
% %
% For training our models, we use all unique fingerprinting function bytecode sequences (797 in total) and set aside $\sim$10\% for our test set.
% %
% We then repeat each fingerprinting function in the training set four times in our final training set.
% %
% We select a random subsample of negative examples such that our training set is $\sim$10\% positive examples, in total our training set was 28,680 examples for functions and 170,155 examples for scripts.
% %
% We use a subset of positive and negative examples for testing, ensuring no function used in training is present in the test set.
\textbf{Dataset Imbalance and Sampling Strategy.}
The final processed dataset exhibits a significant class imbalance, with non-fingerprinting (non-FP) functions vastly outnumbering fingerprinting (FP) functions (4,670 FP vs. approx. 405M non-FP), as shown in Table~\ref{tab:original_dataset_stats}.
This imbalance is anticipated, given that fingerprinting behavior, particularly involving the specific high-entropy APIs targeted by our labeling heuristics, is relatively infrequent compared to the bulk of general web script functionality.
Training directly on such an imbalanced dataset would likely result in a classifier heavily biased towards the majority (non-FP) class, leading to poor detection of the minority (FP) class.

To address this imbalance and create a more effective training set for our supervised models, we employ a combined sampling strategy.
We retain all identified positive examples (the 4,670 unique fingerprinting functions after cleaning and deduplication) to ensure the model learns from all available instances of fingerprinting behavior.
To balance the classes for training, we then randomly undersample the vastly larger set of negative (non-FP) functions.
Specifically, we select approximately 20 times the number of FP functions from the non-FP function pool (resulting in roughly 93,400 non-FP function samples).
This creates the final training dataset with a FP:non-FP ratio of approximately 1:20.
The full, original distribution (after cleaning and deduplication, but before training set balancing) is used for the test set to evaluate the model's performance under realistic conditions reflecting real-world class prevalence.

\subsection{Dataset Analysis}
To understand the characteristics of fingerprinting versus non-fingerprinting scripts and functions in our cleaned dataset, we analyze function counts per script and bytecode sequence lengths.

\begin{figure}[htbp]
    \centering
    \includegraphics[width=0.7\linewidth]{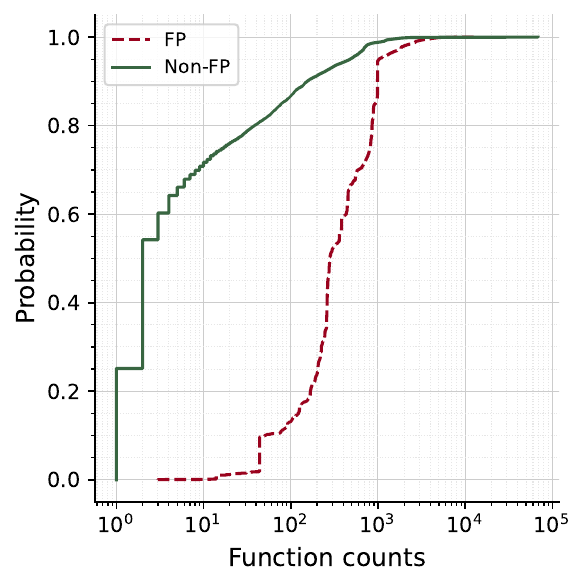}
    \caption{Cumulative distribution of the number of functions in fingerprinting (FP) versus non-fingerprinting (non-FP) scripts.}
    \label{fig:functions_count_fp_vs_benign}
    \vspace{-0.1 in} % Consider removing/adjusting
\end{figure}

\textbf{Functions per Script.} Figure~\ref{fig:functions_count_fp_vs_benign} shows the Cumulative Distribution Function (CDF) of the number of functions per script, comparing scripts labeled as fingerprinting (containing at least one fingerprinting function) versus non-fingerprinting scripts. 
A vast majority (~90\%) of non-fingerprinting scripts contain fewer than 100 functions. 
In contrast, scripts containing fingerprinting functions typically have significantly more functions, with the CDF rising sharply between 100 and 1,000 functions. 
This suggests that fingerprinting logic is often embedded within larger, more complex scripts that likely perform other tasks, reinforcing the need for function-level detection to avoid breaking legitimate functionality contained within the same script.

\textbf{Mixed Scrips.} We investigate the prevalence of mixed scripts in our dataset, where both legitimate functional and fingerprinting scripts are combined within a single script.
Our analysis reveals that all 7,539 fingerprinting scripts identified are mixed scripts.
Surprisingly, not a single script is composed entirely of functions labeled as fingerprinting.
Therefore, blocking entire scripts based solely on the presence of one fingerprinting function would likely result in widespread disruption.

\begin{figure*}[htbp]
    \centering
    % --- Left Figure ---
    % Width is now relative to \textwidth to use the full page width.
    \begin{subfigure}[b]{0.48\textwidth}
        \centering
        \includegraphics[width=0.7\linewidth]{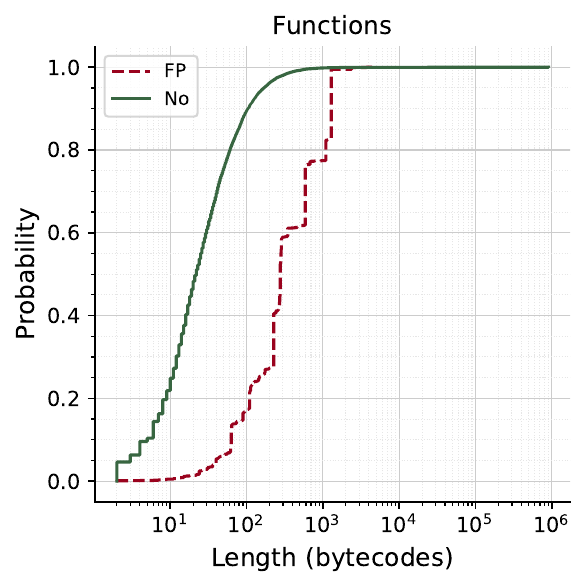}
        \caption{Functions}
        \label{subfig:fingerprinting_functions}
    \end{subfigure}
    \hfill % Add \hfill back to create space between the figures.
    % --- Right Figure ---
    \begin{subfigure}[b]{0.48\textwidth}
        \centering
        \includegraphics[width=0.7\linewidth]{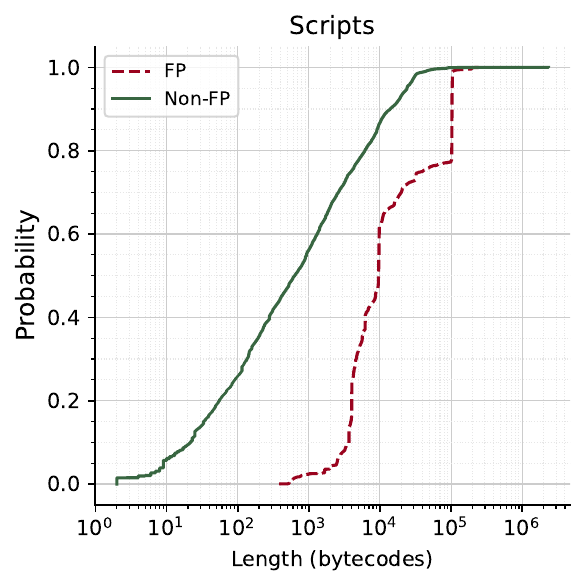}
        \caption{Scripts}
        \label{subfig:benign_scripts}
    \end{subfigure}
    \caption{Comparative distribution of bytecode lengths for functions (a) and scripts (b).}
    \label{fig:bytecode_length}
    \vspace{-0.1in}
\end{figure*}

\textbf{Bytecode Length.} We also analyze the distribution of bytecode sequence lengths (number of bytecode instructions) for individual functions and concatenated script bytecode, as shown in Figure~\ref{fig:bytecode_length}. 
The bytecode sequence of a script is constructed by concatenating the bytecode sequences of individual functions of the script together.
A clear difference emerges: around 90\% of non-fingerprinting functions have fewer than 100 bytecode instructions. 
Conversely, 90\% of fingerprinting functions have bytecode lengths ranging between approximately 100 and 1,000 instructions. 
A similar, though less pronounced, trend holds for entire scripts. 
This indicates that functions performing fingerprinting tend to be computationally more complex or involve more operations than average non-fingerprinting functions, providing a potential signal for classification.

% \begin{figure}[htbp]
%     \centering
%     \begin{subfigure}[b]{0.49\columnwidth}
%         \centering
%         \includegraphics[width=\linewidth]{plots/CDF_bytecode_length_in_functions.pdf}
%         \caption{Functions}
%         \label{subfig:fingerprinting_functions}
%     \end{subfigure}
%     \hfill
%     \begin{subfigure}[b]{0.49\columnwidth}
%         \centering
%         \includegraphics[width=\linewidth]{plots/CDF_bytecode_length_in_scripts_with_non_anonymous_functions.pdf}
%         \caption{Scripts}
%         \label{subfig:benign_scripts}
%     \end{subfigure}
%     \caption{Comparative distribution of bytecode lengths for functions (a) and scripts (b).}
%     \label{fig:bytecode_length}
%     \vspace{-0.1 in}
% \end{figure}

% \subsection{Model Architecture and Training}\label{sec:model_parameters}
\subsection{Implementation Details}\label{sec:model_parameters}

\textbf{Static Embeddings (Word2Vec/FastText).} For baseline models requiring static embeddings, we explored Word2Vec~\cite{mikolov2013efficient} and FastText~\cite{joulin2016bag}. 
Recognizing that bytecode sequences are long but use a limited vocabulary compared to natural language, we prioritized computational efficiency by testing lower embedding dimensions (50, 100, 200) rather than the 300 often used for NLP~\cite{mikolov2013efficient}. 
We experimented with context window sizes (3 and 5) and training epochs (100 and 150) on the bytecode corpus derived from our training split. 
The optimal configuration for Word2Vec, based on downstream performance with the Random Forest classifier, includes an embedding dimension of 100 and a window size of 3. 
For FastText, these include an embedding dimension of 50 and a window size of 3.

\textbf{Random Forest Configuration.} We tuned the Random Forest classifier using GridSearchCV~\cite{pedregosa2011scikit} on the validation set (derived from the balanced training data). 
Tested hyperparameters included split criterion ('gini', 'entropy'), max depth (10, 20, 30), max features ('sqrt', 'log2'), minimum samples per split (5, 10), and the number of estimator (100, 200).
The optimal configuration for this classifier includes the following settings: criterion='entropy', max\_depth=30, max\_features='sqrt', min\_samples\_split=5, and n\_estimators=200.

\textbf{Transformer Architecture.} Our custom Transformer model is implemented using JAX/TensorFlow~\cite{frostig2018compiling, abadi2015tensorflow} and inspired by the original design~\cite{vaswani2017attention} but adapted for bytecode classification. 
We determined the final architecture through experimentation on the validation set. 
The selected model for function classification employs:
\begin{itemize}[noitemsep, topsep=0pt, leftmargin=*]
    \item Input embedding dimension: 256.
    \item Transformer blocks: 1 layer.
    \item Attention heads: 4 heads per layer.
    \item Feed-forward dimension: 512 neurons in the hidden layer of the position-wise feed-forward networks within each block.
    \item Classification head: Global Average Pooling followed by two dense layers with ReLU activation, interspersed with dropout layers (rate 0.1), and a final sigmoid output neuron.
\end{itemize}
The selected model for script classification employs:
\begin{itemize}[noitemsep, topsep=0pt, leftmargin=*]
    \item Input embedding dimension: 128.
    \item 1D Convolutional Layer (Optional): A 1D convolution with kernel size and stride of 2 along the sequence axis, this reduces the sequence length for attention and saves memory.
    \item Transformer blocks: 1 layer.
    \item Attention heads: 4 heads per layer.
    \item Feed-forward dimension: 256 neurons in the hidden layer of the position-wise feed-forward networks within each block.
    \item Classification head: Same as function model.
\end{itemize}
These configurations provided the best trade-off between performance and complexity in our experiments.

\textbf{Transformer Training Details:} For function classification, the Transformer model was trained for 16 epochs on the balanced training set using the Adam optimizer~\cite{kingma2014adam} with binary cross-entropy loss.
We employed a batch size of 128.

For script classification, which has more memory requirements due to the longer sequence length, we employed a batch size of 8 and 16.
We trained the model for 10 epochs, since the script classification dataset was larger.
We found these hyperparameters empirically to offer a good balance between model complexity and generalization based on validation performance~\cite{srivastava2014dropout}.

\section{Evaluation}\label{sec:evaluation}

In this section, we evaluate the effectiveness of \name{} in detecting fingerprinting functions at the JavaScript function level, and its overhead when browsing the web. 
We assess our primary Transformer-based classifier and compare it against a Random Forest baseline, using the dataset and methodologies detailed in Sections~\ref{sec:experimental_setup} and~\ref{sec:\name}. 
Specifically, we compare the models according to their accuracy, precision, and recall. 
Furthermore, we evaluate \name{}'s script-level robustness against code obfuscation.

\subsection{Function-Level Classification Performance}\label{subsec:fn_level_eval}

\textbf{Experimental Setup:} We used the processed dataset described in Section~\ref{sec:experimental_setup}, derived from crawling 100k websites. 
The data was split into training (90\%) and testing (10\%) sets, ensuring no duplicate function bytecode sequences were found across splits. 
Due to class imbalance, the training set was balanced by undersampling the non-fingerprinting class to a 1:10 ratio (FP:non-FP) after oversampling positive examples. 
For the Random Forest baseline, Word2Vec and FastText embeddings were trained on the training set's bytecode corpus, and the resulting averaged function vectors were used for training the RF model. 
For the Transformer classifier, embeddings were learned end-to-end during training on the bytecode sequences.

\textbf{Results.} Table~\ref{tab:performance} presents the core classification results on the test set. 
Our Transformer classifier significantly outperforms the Random Forest baselines across most metrics, achieving an accuracy of 98.9\%, precision of 84.0\%, and recall of 85.1\%. 
While Random Forest with FastText embeddings achieves high precision (94.8\%), its recall is considerably lower (66.3\%), indicating it misses a substantial portion of fingerprinting functions. Word2Vec embeddings yield even lower performance for the Random Forest model. 
The Transformer's superior performance, especially in recall, highlights the benefit of processing the full bytecode sequence and leveraging contextual embeddings learned end-to-end, compared to using averaged static embeddings which lose sequential information.

The choice between precision and recall depends on the deployment goal. High recall, as achieved by the Transformer, is crucial if the priority is to minimize missed fingerprinting instances (false negatives), even at the cost of some false positives. 
High precision, where RF+FastText excels, is important if minimizing disruption to non-fingerprinting functions (false positives) is paramount. 
The high overall accuracy of the Transformer suggests it provides the best balance for effectively detecting the majority of fingerprinting functions present in the wild.

\begin{table}[htbp]
\centering
\caption{Model performance for function classification using different classifiers and embeddings on the test set.}
\label{tab:performance}
\small % Use a standard font size command for readability
% The key change is making the numeric columns flexible using the 'X' type.
% This prevents the 'Embedding Model' column from overlapping them.
\begin{tabularx}{\columnwidth}{@{} l >{\RaggedRight}X *{5}{>{\centering\arraybackslash}X} @{}}
\toprule
\textbf{Classifier} & \textbf{Embed. Model} & \textbf{\shortstack{Acc.\\(\%)}} & \textbf{\shortstack{Prec.\\(\%)}} & \textbf{\shortstack{Recall\\(\%)}} & \textbf{\shortstack{ROC\\AUC (\%)}} & \textbf{\shortstack{PR\\AUC (\%)}} \\
\midrule
Transformer & Transf. (learned) & 98.9 & 84.0 & 85.1 & 93.3 & 81.6 \\
\midrule
\multirow{2}{*}{Random Forest} & Word2Vec & 93.6 & 80.0 & 60.0 & 79.0 & 60.7 \\
& FastText & 94.8 & 94.8 & 66.3 & 83.1 & 67.6 \\
\bottomrule
\end{tabularx}
\vspace{-0.1in}
\end{table}

% \begin{table}[htbp] % Use htbp for better placement
% \centering
% \caption{Model performance for function classification using different classifiers and embeddings on the test set.}
% \label{tab:performance}
% \resizebox{\columnwidth}{!}{
% \begin{tabular}{@{}l|lccccc@{}}
% \toprule
% \multicolumn{1}{c|}{\textbf{Classifier}} & \multicolumn{1}{c}{\textbf{Embedding Model}} & \textbf{Accuracy(\%)} & \textbf{Precision(\%)} & \textbf{Recall(\%)} & \textbf{ROC AUC (\%)} & \textbf{PR AUC (\%)} \\ 
% \midrule 
% Transformer                    & Transformer (learned) & 98.9 & 84.0 & 85.1 & 93.3 & 81.6\\
% \midrule
% \multirow{2}{*}{Random Forest} & Word2Vec   & 93.6 & 80.0 & 60.0 & 79.0 & 60.7 \\
%                               & FastText   & 94.8 & 94.8 & 66.3 & 83.1 & 67.6 \\ 

% \bottomrule
% \end{tabular}
% }
% \vspace{-0.1in} 
% \end{table}

\subsection{Comparison to State-of-the-Art (Script-Level)}\label{subsec:script_level_eval}

To compare \name{} with prior work that operates at the script-level, we adapted our approach. 
As existing methods primarily detect fingerprinting based on entire scripts~\cite{iqbal2021fingerprinting, rizzo2018machine, van2018detection}, we created a script-level version of our dataset by concatenating the bytecode of all functions within a script and labeling the script as FP if it contained at least one fingerprinting function. 
We then trained our Transformer architecture on this script-level data.

We compare this against a reimplementation of the static AST-based approach, representative of methods used in studies by Iqbal et al.~\cite{iqbal2021fingerprinting}, Rizzo~\cite{rizzo2018machine}, and van Zalingen and Haanen~\cite{van2018detection}. 
Reconstructing this model involved using the script source code collected during our crawls (Section~\ref{sec:data_collection}) and employing the Esprima library's `parseScript' function~\cite{esprima_parsing} to convert each script's source code into its corresponding AST. 
We then traversed these ASTs hierarchically, segmenting them into pairs of parent and child nodes. 
In this representation, parent nodes capture contextual structures (such as `for' loops, `try' statements, or `if' conditions), while child nodes represent the specific functions or operations executed within those contexts (for instance, API calls such as \texttt{createElement}, \texttt{toDataURL}, or \texttt{measureText}). 
To create input vectors for classification, we encoded the presence of these parent-child pairs using one-hot vectors, where each vector position corresponds to a unique potential parent-child combination observed across the dataset. 
Within a script's vector, a value of `1' at a specific position indicates that the corresponding parent-child combination exists in its AST, while a `0' signifies its absence. 
Finally, following the methodology described for the static component in FP-Inspector~\cite{iqbal2021fingerprinting}, we used these one-hot feature vectors representing the script's AST patterns as input to train a standard decision tree classifier.

Table~\ref{tab:comparison} shows the results of this script-level comparison using the test set derived from our dataset. 
\name{}'s Transformer, adapted to the script-level, demonstrates exceptionally high performance across all metrics: 99.7\% accuracy, 92.1\% precision, and 96.9\% recall. 
The reimplemented AST-based classifier achieves considerably lower performance, with 97.5\% accuracy, 85.0\% precision, and only 80.0\% recall. 
The substantial difference, particularly the nearly 17\% higher recall achieved by \name{}, strongly underscores the superiority of the bytecode-based approach compared to AST analysis for script-level fingerprinting detection. 
Bytecode captures semantics closer to execution and is less susceptible to syntactic variations and obfuscation, leading to significantly better identification of true positive fingerprinting scripts.

\begin{table}[htbp]
\centering
\caption{Comparative performance of \name{} (Transformer, script-level) and representative AST-based classifiers in script classification.}
\label{tab:comparison}
\small
% The key change is in the column specification below.
\begin{tabularx}{\columnwidth}{@{} >{\RaggedRight\arraybackslash}X ccccc @{}}
\toprule
\textbf{Method} & \textbf{\shortstack{Acc.\\(\%)}} & \textbf{\shortstack{Prec.\\(\%)}} & \textbf{\shortstack{Recall\\(\%)}} & \textbf{\shortstack{ROC\\AUC (\%)}} & \textbf{\shortstack{PR\\AUC (\%)}} \\
\midrule
\name{} (Transformer, script-level) & 99.7 & 92.1 & 96.9 & 99.5 & 92.3 \\
\midrule
AST-based classifiers from prior work (Random-Forest, script-level) & 97.5 & 85.0 & 80.0 & 90.0 & 68.0 \\
\bottomrule
\end{tabularx}
\vspace{-0.1in}
\end{table}

% \begin{table}[htbp]
% \centering
% \caption{Comparative performance of \name{} (Transformer, script-level) and representative AST-based classifiers in script classification.}
% \label{tab:comparison}
% \resizebox{\columnwidth}{!}{% 
% % Define columns with vertical line after first column
% \begin{tabular}{@{}l|ccccc@{}} 
% \toprule
% % Header: Use multicolumn for first header cell to maintain '|' and center 'Method'
% \multicolumn{1}{c|}{\textbf{Method}} & \textbf{Accuracy (\%)} & \textbf{Precision (\%)} & \textbf{Recall (\%)} & \textbf{ROC AUC (\%)} & \textbf{PR AUC (\%)}  \\ 
% \midrule 
% % Data Rows: Use nested tabular for multi-line content in the first column
% % The outer 'l' ensures the nested tabular is left-aligned in the cell
% \begin{tabular}[t]{@{}l@{}} % [t] aligns the top of this tabular with the baseline
%   \name \\ (Transformer, Script-Level) 
% \end{tabular}
%     & 99.7                   & 92.1                    & 96.9           & 99.5          & 92.3       \\ 
% \midrule 
% \begin{tabular}[t]{@{}l@{}} % [t] aligns the top
%   AST-based classifiers~\cite{iqbal2021fingerprinting, rizzo2018machine, van2018detection}\\ (RandomForest, script-level) 
% \end{tabular}
%     & 97.5                   & 85.0                    & 80.0           & 90.0       & 68.0      \\ 
% \bottomrule 
% \end{tabular}%
% } % Closing brace for resizebox
% \vspace{-0.1in} 
% \end{table}

\subsection{Robustness Evaluation}\label{sec:robustness}

A key advantage claimed for bytecode analysis is its robustness against common evasion techniques. 
We evaluate this against URL manipulation and code obfuscation.

\textbf{Robustness against URL manipulations.} As \name{} analyzes function bytecode content directly and does not rely on the script's origin URL or domain for its classification decision (unlike filter lists or some features in~\cite{iqbal2021fingerprinting}), it is inherently robust against URL-based evasion techniques such as CNAME cloaking~\cite{Dimova2021TheCO, dao2021cname} or path/parameter randomization~\cite{wang2016webranz}. 
This allows \name{} to detect fingerprinting behavior regardless of how or where the script is hosted.

\textbf{Robustness against code obfuscation.} Fingerprinting scripts often employ obfuscation techniques to complicate static inspection and evade detection~\cite{skolka2019anything}.
As we discussed in Section~\ref{sec:v8}, our model operates on bytecode sequences that retain only the opcode instructions, omitting auxiliary details such as memory addresses, operands, and bytecode offsets.
This abstraction aligns with our goal of building a generalizable and obfuscation-resilient detection approach.
Common obfuscation strategies—such as string encoding, variable renaming, and control flow flattening—primarily target operands and syntactic constructs, while typically preserving the underlying operational behavior reflected in opcode instructions~\cite{skolka2019anything}.
At the function-level, V8 bytecode is relatively robust to obfuscation, since the compiled bytecode retains semantic structure even when the source code is heavily transformed. 
However, at the script-level—where bytecode sequences represent full-page execution contexts—robustness degrades due to non-deterministic factors such as script loading order, lazy compilation, and network timing~\cite{cabrera2019scalable}.
These factors can cause significant variation even across semantically identical scripts.

To evaluate the impact of obfuscation at the script level, we assess \name{}'s performance---which is trained on real-world scripts consisting of a mixture of unobfuscated and obfuscated code (Section~\ref{subsec:script_level_eval})---on re-obfuscated scripts.
{\sloppy To generate obfuscated versions of collected scripts, we use two distinct and publicly available tools: JavaScript-Obfuscator~\cite{js_obfuscator}, which applies heavy transformations such as variable renaming, string encoding, and control flow flattening; and the Google Closure Compiler~\cite{google_closure}, which focuses on code optimization through renaming and simplification.
We then use our instrumented V8 engine (Section~\ref{sec:v8}) to extract bytecode sequences from these obfuscated scripts.
The evaluation includes 47,000 JavaScript-Obfuscator samples and 11,000 Google Closure obfuscated samples, each preserving a 1:20 ratio of fingerprinting to non-fingerprinting scripts.
{\sloppy As shown in the first row of Table~\ref{tab:obfuscation}, \name{}, when trained only on real-world scripts, performs poorly on re-obfuscated inputs.
For example, when obfuscating with JavaScript Obfuscator, precision remains moderate (60.5\%), but recall falls to 0.1\%, indicating a near-total failure to detect fingerprinting scripts. 
A similar trend is observed with the Google Closure Compiler, where recall is only 2.8\%.\par}

To mitigate this vulnerability, we train \name{}'s script-level model on a an augmented dataset that combines real-world and re-obfuscated scripts.
Our final training set consists of approximately 100,000 scripts with a 1:20 fingerprinting-to-non-fingerprinting ratio, augmented with the aforementioned obfuscated samples.
The model is then evaluated on a balanced set of previously unseen real-world and re-obfuscated samples. 
As shown in the last two rows of Table~\ref{tab:obfuscation}, augmented \name{} significantly improves generalization.
For instance, recall jumps from 0.1\% to 92.1\% with JavaScript Obfuscator and from 2.8\% to 78.0\% with Google Closure. 
These results demonstrate that exposing the model to diverse obfuscation styles during training enables it to generalize effectively and detect fingerprinting behavior in obfuscated scripts.\par}

\begin{table}[htbp]
\centering
\caption{Evaluation of script-level \name{} and its augmented variant on scripts obfuscated using various techniques.}
\label{tab:obfuscation}
\small
% Use tabularx with two wrapping columns for text and three compact columns for numbers.
\begin{tabularx}{\columnwidth}{@{} >{\RaggedRight}X >{\RaggedRight}X ccc @{}}
\toprule
\textbf{Classifier} & \textbf{Obfuscator} & \textbf{\shortstack{Acc.\\(\%)}} & \textbf{\shortstack{Prec.\\(\%)}} & \textbf{\shortstack{Recall\\(\%)}} \\
\midrule
% \multirow is used to merge rows for the classifier groups.
\multirow{2}{*}{\name} & JavaScript Obfuscator & 55.0 & 60.5 & 0.1  \\
                       & Google Closure        & 56.8 & 93.8 & 2.8  \\
\midrule
% \shortstack is a simple way to create multi-line text inside a cell.
\multirow{2}{*}{\shortstack{Augmented\\\name}} & JavaScript Obfuscator & 89.1 & 85.0 & 92.1 \\
                                              & Google Closure        & 89.9 & 82.9 & 78.0 \\
\bottomrule
\end{tabularx}
\vspace{-0.1in}
\end{table}

% \begin{table}[htbp]
% \centering
% % Removed \large
% \caption{Evaluation of script-level \name{} and its augmented variant on scripts obfuscated using various techniques.}
% \label{tab:obfuscation}
% \resizebox{\columnwidth}{!}{%
% \begin{tabular}{@{}llccc@{}}
% \toprule
% % Use parbox for the header spanning two lines
% \textbf{\parbox{3cm}{\centering Script-level\\Classifier}} & % Adjusted width to 3cm, same as row labels
%   \multicolumn{1}{c}{\textbf{Obfuscator}} &
%   \multicolumn{1}{l}{\textbf{Accuracy(\%)}} &
%   \multicolumn{1}{l}{\textbf{Precision(\%)}} &
%   \multicolumn{1}{l}{\textbf{Recall(\%)}} \\ 
% \midrule
% % Use \parbox for multi-line content and centering within the \multirow cell
% \multirow{2}{*}{\parbox{3cm}{\centering \name}} & % Added centering
%                       JavaScript Obfuscator & 55.0 & 60.5 & 0.1  \\
%                       & Google Closure        & 56.8 & 93.8 & 2.8  \\ 
% \midrule
% % Use \parbox for multi-line content and centering within the \multirow cell
% \multirow{2}{*}{\parbox{3cm}{\centering Augmented\\\name}} & 
%                       JavaScript Obfuscator & 89.1 & 85.0 & 92.1 \\ 
%                       & Google Closure        & 89.9 & 82.9 & 78.0 \\ 
% \bottomrule
% \end{tabular}%
% }
% \vspace{-0.1 in}
% \end{table}

\subsection{Real-Time Detection Overhead in the Browser}
Next, we assess the runtime overhead introduced by enabling on-device signature matching against JavaScript function bytecodes.

To detect fingerprinting behaviors, we first generate a list of non-cryptographic hashes corresponding to known fingerprinting function bytecodes and embed these signatures into the V8 engine source code. 
Within \texttt{interpreter.cc}, we call the \texttt{Disassembly()} method from the \texttt{BytecodeArray} class to translate raw binary bytecode into its symbolic instruction format. 
To manage the verbosity of the output, we apply the same selective instrumentation approach detailed in Section~\ref{sec:v8}. 
We further modify V8 to compute the hash of each function's bytecode sequence at runtime and compare it against the embedded signature list to detect fingerprinting functions during execution.

\begin{figure}[htbp]
    \centering
    \includegraphics[width=0.8\linewidth]{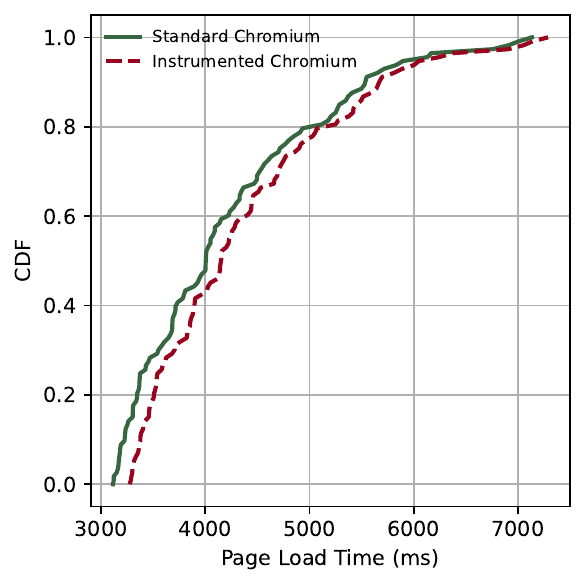}
    \caption{CDF plot comparing page load times of the baseline (standard Chromium) and our instrumented Chromium with signature matching, across 1,000 websites.}
    \label{fig:cdf_overhead}
    \vspace{-0.1 in} % Consider removing/adjusting
\end{figure}

To quantify the overhead, we crawl 1,000 websites using our modified Chromium browser with lightweight signature matching enabled and compare its performance against an unmodified baseline Chromium build. 
Figure~\ref{fig:cdf_overhead} illustrates the distribution of page load times for both the baseline and instrumented Chromium across 1,000 websites.
As shown, the CDF curves for the two versions are closely aligned.
Our measurements show that the signature matching mechanism introduces an average additional latency of 158.74ms, which represents only a 4\% increase of the overall median page load time in our sample.
The distribution of overhead is tightly bounded, with the 5th percentile at 117.54ms, 25th percentile at 146.22ms, median at 158.62ms, 75th percentile at 170.30ms, and 95th percentile at 199.21ms.
These results demonstrate that the runtime cost of on-device bytecode matching remains consistently low, introducing negligible impact on overall browser performance while preventing fingerprinting functions---and \emph{not} entire scripts---from being executed.

\section{Limitations and Future Work}\label{sec:limitations}

While \name{} demonstrates strong performance and robustness against common evasion techniques, we acknowledge several limitations inherent in our current methodology and dataset construction, alongside potential avenues for future research.

\subsection{Impact of Anonymous Functions and Scripts}
Our approach to generating ground truth labels necessitates mapping function execution traces to their corresponding bytecode sequences using script URL, script ID, and function name (Section~\ref{sec:assiging_lables}). 
This requirement presented a significant challenge, as anonymous functions (lacking names) or scripts without canonical URLs (e.g., those loaded via `eval` or complex dynamic means) could not be reliably mapped and were consequently excluded during the data cleaning phase (Section~\ref{subsec:dataset_construction}). 
As shown in Table~\ref{tab:original_dataset_stats}, this filtering step removed a substantial number of functions, including a large fraction of potential fingerprinting instances initially identified by high-entropy API usage. 
This necessary exclusion limits the diversity of functions represented in our training set, potentially affecting the model's ability to generalize to websites heavily reliant on anonymous functions or unconventional script loading patterns.

However, it is important to clarify the scope of this limitation. 
It primarily affects the generation of labeled training data. 
The \name{} Transformer model itself, once trained, analyzes bytecode structure and does not inherently require function names or script URLs for classification. It learns bytecode patterns associated with fingerprinting. 
Therefore, the model may still successfully classify anonymous functions or inline scripts during deployment if their bytecode exhibits patterns similar to those learned from the named functions in the training set.
Techniques like code inlining (embedding scripts directly in HTML) are not necessarily an evasion strategy against \name's core detection capability, as the V8 engine still generates bytecode for these inline script blocks which can then be analyzed. 
\revedit{However, we acknowledge that \name’s function-level classifier has not been explicitly evaluated on anonymous or eval-loaded functions, and its generalization to such cases remains unverified.}

Future work could focus on developing alternative ground truth generation methods less reliant on explicit identifiers, perhaps incorporating dynamic analysis techniques for associating traces with anonymous functions or employing code similarity metrics on bytecode to propagate labels.

\subsection{Reliance on Heuristic-Based Ground Truth}
We established the ground truth for this study using heuristics based on well-documented fingerprinting techniques involving specific high-entropy API call patterns (Section~\ref{sec:Detecting_fingerprinting_functions}), drawing from established research~\cite{iqbal2021fingerprinting, englehardt2016online}. 
This approach was necessary due to the infeasibility of manually labeling the millions of functions collected, and it provides a high-precision dataset for training, aligning with practices in related work~\cite{iqbal2021fingerprinting, englehardt2016online}. 
Our evaluation (Section~\ref{sec:evaluation}) confirms that models trained on this data achieve high accuracy for known fingerprinting methods. 
However, this reliance on predefined heuristics \revedit{introduces an inherent bias: the heuristics are tailored to known fingerprinting techniques and prioritize precision. As a result, they may fail to capture novel or evolving fingerprinting behaviors that deviate from documented patterns or utilize different API sequences}~\cite{bahrami2021fp}.
% \name, in its current form, may not detect entirely novel fingerprinting techniques that exploit different APIs or interaction patterns not covered by our heuristics~\cite{bahrami2021fp}. 
Addressing this requires ongoing research. 
Exploring semi-supervised or unsupervised learning methods capable of identifying anomalous or suspicious bytecode patterns directly, without depending on predefined API lists, could enable the detection of emerging fingerprinting strategies and represents a promising direction for future work.

\subsection{Browser and Engine Dependency}
\name's current implementation is specifically tailored to Chromium's V8 engine. 
We leverage V8's bytecode instruction set as the feature representation for our classifiers. 
\revedit{Because JavaScript bytecode is not standardized across different browser engines (such as SpiderMonkey in Firefox or JavaScriptCore in Safari), each having its own distinct instruction set and compilation pipeline, a model trained on V8 bytecode is unlikely to function correctly on bytecode generated by other engines\emph{without significant adaptation}. 
Extending \name's approach to achieve cross-browser compatibility would necessitate instrumenting and \emph{collecting data from each target engine and training engine-specific models}.}

Furthermore, bytecode instruction sets can evolve even within the same engine (V8) across different versions, driven by new JavaScript language features or internal optimizations. 
This implies that maintaining optimal performance for \name over time may require periodic retraining using bytecode collected from updated V8 versions. 
While the core methodology remains valid, this potential need for retraining represents an ongoing maintenance consideration. 

Future research should explore techniques for cross-engine bytecode translation or the identification of higher-level, potentially engine-invariant features derivable from bytecode to enhance model portability across browsers and versions.

\section{Conclusion}\label{sec:conclusion}
Browser fingerprinting presents a persistent and evolving challenge to user privacy, leveraging subtle browser and device characteristics to enable cross-site tracking often immune to conventional defenses. Existing countermeasures frequently fall short, either lacking precision and causing website breakage (e.g., script-level blocking) or proving vulnerable to common evasion tactics like code obfuscation and URL manipulation (e.g., filter lists, AST-based analysis). Addressing this requires a robust, precise, and proactive detection mechanism.

In this paper, we introduced \name{}, the first system to utilize V8 engine bytecode specifically for detecting fingerprinting behaviors at the JavaScript function level. By analyzing the intrinsic bytecode patterns of individual functions during compilation, \name{} achieves fine-grained detection before execution occurs. We demonstrated the efficacy of a Transformer-based classifier trained on function-level bytecode sequences derived from a large-scale crawl of 100k websites, where ground truth was established using heuristic analysis of execution traces.

Our evaluation confirms the advantages of this approach. \name{} achieves high detection accuracy (99.7\% at the script level), precision (92.1\%), and recall (96.9\%), significantly outperforming representative AST-based methods, especially demonstrating superior robustness against obfuscated JavaScript code. Its function-level granularity is crucial, given our finding that fingerprinting scripts are typically mixed-purpose, mitigating the web breakage associated with coarser script-level blocking. We also demonstrated the feasibility of efficient on-device deployment through preliminary measurements showing that lightweight signature matching introduces only negligible (average 4\%) page load latency.

\name{} offers a practical and effective framework for fingerprinting mitigation. It moves beyond the limitations of existing methods by providing a robust, precise, function-level analysis grounded in the code representation used by the JavaScript engine itself. By enabling targeted, pre-execution intervention, it enhances user privacy with minimal impact on web compatibility. Future work includes refining signature generation for minimal overhead and exploring cross-browser compatibility.

\bibliographystyle{ACM-Reference-Format}
\bibliography{sections/ref}

\appendix
\section{List of high entropy APIs}\label{sec:appendix1}
The following table lists 322 features, consisting exclusively of API call counts that have been flagged as ‘high entropy’ by Chromium.

\onecolumn
\begin{footnotesize}
\setlength{\tabcolsep}{1pt}
\begin{longtable}{p{0.3\textwidth} p{0.3\textwidth} p{0.31\textwidth}}
\caption{List of APIs in Chrome explicitly flagged by Chromium as ‘High Entropy APIs’.}
\label{tab:highEntropyAPIs} \\
\toprule
\multicolumn{3}{c}{\textbf{High Entropy APIs}} \\
\midrule
\endfirsthead

\toprule
\multicolumn{3}{c}{\textbf{High Entropy APIs (continued)}} \\
\midrule
\endhead

\bottomrule
\endfoot

\bottomrule
\endlastfoot
\multicolumn{1}{|l|}{AnalyserNode.getByteFrequencyData}                     & \multicolumn{1}{l|}{NavigatorUAData.platform.get}                                   & RTCIceCandidate.candidate.get                     \\* \midrule
\multicolumn{1}{|l|}{AnalyserNode.getByteTimeDomainData}                    & \multicolumn{1}{l|}{NavigatorUAData.toJSON}                                         & RTCIceCandidate.port.get                          \\* \midrule
\multicolumn{1}{|l|}{AnalyserNode.getFloatFrequencyData}                    & \multicolumn{1}{l|}{Navigator.userAgent.get}                                        & RTCIceCandidate.relatedAddress.get                \\* \midrule
\multicolumn{1}{|l|}{AnalyserNode.getFloatTimeDomainData}                   & \multicolumn{1}{l|}{Navigator.vendor.get}                                           & RTCIceCandidate.relatedPort.get                   \\* \midrule
\multicolumn{1}{|l|}{AudioBuffer.copyFromChannel}                           & \multicolumn{1}{l|}{Navigator.vendorSub.get}                                        & RTCRtpReceiver.getCapabilities                    \\* \midrule
\multicolumn{1}{|l|}{AudioBuffer.getChannelData}                            & \multicolumn{1}{l|}{Navigator.webkitGetUserMedia}                                   & RTCRtpSender.getCapabilities                      \\* \midrule
\multicolumn{1}{|l|}{AudioContext.baseLatency.get}                          & \multicolumn{1}{l|}{NetworkInformation.downlink.get}                                & Screen.availHeight.get                            \\* \midrule
\multicolumn{1}{|l|}{AudioContext.constructor}                              & \multicolumn{1}{l|}{NetworkInformation.downlinkMax.get}                             & Screen.availLeft.get                              \\* \midrule
\multicolumn{1}{|l|}{AudioContext.outputLatency.get}                        & \multicolumn{1}{l|}{NetworkInformation.effectiveType.get}                           & Screen.availTop.get                               \\* \midrule
\multicolumn{1}{|l|}{AudioNode.connect}                                     & \multicolumn{1}{l|}{NetworkInformation.rtt.get}                                     & Screen.availWidth.get                             \\* \midrule
\multicolumn{1}{|l|}{AuthenticatorAttestationResponse.getTransports}        & \multicolumn{1}{l|}{NetworkInformation.saveData.get}                                & Screen.colorDepth.get                             \\* \midrule
\multicolumn{1}{|l|}{BackgroundFetchRegistration.failureReason.get}         & \multicolumn{1}{l|}{NetworkInformation.type.get}                                    & ScreenDetailed.devicePixelRatio.get               \\* \midrule
\multicolumn{1}{|l|}{BaseAudioContext.createDynamicsCompressor}             & \multicolumn{1}{l|}{OfflineAudioContext.constructor}                                & ScreenDetailed.isInternal.get                     \\* \midrule
\multicolumn{1}{|l|}{BaseAudioContext.createOscillator}                     & \multicolumn{1}{l|}{OfflineAudioContext.startRendering}                             & ScreenDetailed.isPrimary.get                      \\* \midrule
\multicolumn{1}{|l|}{BaseAudioContext.sampleRate.get}                       & \multicolumn{1}{l|}{OffscreenCanvas.convertToBlob}                                  & ScreenDetailed.label.get                          \\* \midrule
\multicolumn{1}{|l|}{BatteryManager.charging.get}                           & \multicolumn{1}{l|}{OffscreenCanvasRenderingContext2D.arc}                          & ScreenDetailed.left.get                           \\* \midrule
\multicolumn{1}{|l|}{BatteryManager.chargingTime.get}                       & \multicolumn{1}{l|}{OffscreenCanvasRenderingContext2D.beginPath}                    & ScreenDetailed.top.get                            \\* \midrule
\multicolumn{1}{|l|}{BatteryManager.dischargingTime.get}                    & \multicolumn{1}{l|}{OffscreenCanvasRenderingContext2D.clearRect}                    & ScreenDetails.oncurrentscreenchange.get           \\* \midrule
\multicolumn{1}{|l|}{BatteryManager.level.get}                              & \multicolumn{1}{l|}{OffscreenCanvasRenderingContext2D.clip}                         & ScreenDetails.oncurrentscreenchange.set           \\* \midrule
\multicolumn{1}{|l|}{BeforeInstallPromptEvent.platforms.get}                & \multicolumn{1}{l|}{OffscreenCanvasRenderingContext2D.closePath}                    & ScreenDetails.onscreenschange.get                 \\* \midrule
\multicolumn{1}{|l|}{BluetoothAdvertisingEvent.appearance.get}              & \multicolumn{1}{l|}{OffscreenCanvasRenderingContext2D.drawMesh}                     & ScreenDetails.onscreenschange.set                 \\* \midrule
\multicolumn{1}{|l|}{BluetoothAdvertisingEvent.name.get}                    & \multicolumn{1}{l|}{OffscreenCanvasRenderingContext2D.ellipse}                      & Screen.height.get                                 \\* \midrule
\multicolumn{1}{|l|}{BluetoothAdvertisingEvent.txPower.get}                 & \multicolumn{1}{l|}{OffscreenCanvasRenderingContext2D.fill}                         & Screen.isExtended.get                             \\* \midrule
\multicolumn{1}{|l|}{BluetoothDevice.name.get}                              & \multicolumn{1}{l|}{OffscreenCanvasRenderingContext2D.fillRect}                     & Screen.onchange.get                               \\* \midrule
\multicolumn{1}{|l|}{CanvasRenderingContext2D.arc}                          & \multicolumn{1}{l|}{OffscreenCanvasRenderingContext2D.fillStyle.get}                & Screen.onchange.set                               \\* \midrule
\multicolumn{1}{|l|}{CanvasRenderingContext2D.beginPath}                    & \multicolumn{1}{l|}{OffscreenCanvasRenderingContext2D.fillStyle.set}                & ScreenOrientation.angle.get                       \\* \midrule
\multicolumn{1}{|l|}{CanvasRenderingContext2D.clearRect}                    & \multicolumn{1}{l|}{OffscreenCanvasRenderingContext2D.fillText}                     & ScreenOrientation.type.get                        \\* \midrule
\multicolumn{1}{|l|}{CanvasRenderingContext2D.closePath}                    & \multicolumn{1}{l|}{OffscreenCanvasRenderingContext2D.font.get}                     & Screen.pixelDepth.get                             \\* \midrule
\multicolumn{1}{|l|}{CanvasRenderingContext2D.drawMesh}                     & \multicolumn{1}{l|}{OffscreenCanvasRenderingContext2D.font.set}                     & Screen.width.get                                  \\* \midrule
\multicolumn{1}{|l|}{CanvasRenderingContext2D.ellipse}                      & \multicolumn{1}{l|}{OffscreenCanvasRenderingContext2D.getImageData}                 & SVGAnimationElement.requiredExtensions.get        \\* \midrule
\multicolumn{1}{|l|}{CanvasRenderingContext2D.fill}                         & \multicolumn{1}{l|}{OffscreenCanvasRenderingContext2D.globalCompositeOperation.get} & SVGAnimationElement.systemLanguage.get            \\* \midrule
\multicolumn{1}{|l|}{CanvasRenderingContext2D.fillRect}                     & \multicolumn{1}{l|}{OffscreenCanvasRenderingContext2D.globalCompositeOperation.set} & SVGGeometryElement.getPointAtLength               \\* \midrule
\multicolumn{1}{|l|}{CanvasRenderingContext2D.fillStyle.get}                & \multicolumn{1}{l|}{OffscreenCanvasRenderingContext2D.isPointInPath}                & SVGGeometryElement.getTotalLength                 \\* \midrule
\multicolumn{1}{|l|}{CanvasRenderingContext2D.fillStyle.set}                & \multicolumn{1}{l|}{OffscreenCanvasRenderingContext2D.isPointInStroke}              & SVGGeometryElement.isPointInFill                  \\* \midrule
\multicolumn{1}{|l|}{CanvasRenderingContext2D.fillText}                     & \multicolumn{1}{l|}{OffscreenCanvasRenderingContext2D.lineTo}                       & SVGGeometryElement.isPointInStroke                \\* \midrule
\multicolumn{1}{|l|}{CanvasRenderingContext2D.font.get}                     & \multicolumn{1}{l|}{OffscreenCanvasRenderingContext2D.measureText}                  & SVGGraphicsElement.requiredExtensions.get         \\* \midrule
\multicolumn{1}{|l|}{CanvasRenderingContext2D.font.set}                     & \multicolumn{1}{l|}{OffscreenCanvasRenderingContext2D.moveTo}                       & SVGGraphicsElement.systemLanguage.get             \\* \midrule
\multicolumn{1}{|l|}{CanvasRenderingContext2D.getImageData}                 & \multicolumn{1}{l|}{OffscreenCanvasRenderingContext2D.rect}                         & SVGMaskElement.requiredExtensions.get             \\* \midrule
\multicolumn{1}{|l|}{CanvasRenderingContext2D.globalCompositeOperation.get} & \multicolumn{1}{l|}{OffscreenCanvasRenderingContext2D.rotate}                       & SVGMaskElement.systemLanguage.get                 \\* \midrule
\multicolumn{1}{|l|}{CanvasRenderingContext2D.globalCompositeOperation.set} & \multicolumn{1}{l|}{OffscreenCanvasRenderingContext2D.roundRect}                    & SVGPatternElement.requiredExtensions.get          \\* \midrule
\multicolumn{1}{|l|}{CanvasRenderingContext2D.isPointInPath}                & \multicolumn{1}{l|}{OffscreenCanvasRenderingContext2D.scale}                        & SVGPatternElement.systemLanguage.get              \\* \midrule
\multicolumn{1}{|l|}{CanvasRenderingContext2D.isPointInStroke}              & \multicolumn{1}{l|}{OffscreenCanvasRenderingContext2D.setTransform}                 & SVGTextContentElement.getComputedTextLength       \\* \midrule
\multicolumn{1}{|l|}{CanvasRenderingContext2D.lineTo}                       & \multicolumn{1}{l|}{OffscreenCanvasRenderingContext2D.shadowBlur.get}               & SVGTextContentElement.getEndPositionOfChar        \\* \midrule
\multicolumn{1}{|l|}{CanvasRenderingContext2D.measureText}                  & \multicolumn{1}{l|}{OffscreenCanvasRenderingContext2D.shadowBlur.set}               & SVGTextContentElement.getExtentOfChar             \\* \midrule
\multicolumn{1}{|l|}{CanvasRenderingContext2D.moveTo}                       & \multicolumn{1}{l|}{OffscreenCanvasRenderingContext2D.shadowColor.get}              & SVGTextContentElement.getStartPositionOfChar      \\* \midrule
\multicolumn{1}{|l|}{CanvasRenderingContext2D.rect}                         & \multicolumn{1}{l|}{OffscreenCanvasRenderingContext2D.shadowColor.set}              & SVGTextContentElement.getSubStringLength          \\* \midrule
\multicolumn{1}{|l|}{CanvasRenderingContext2D.rotate}                       & \multicolumn{1}{l|}{OffscreenCanvasRenderingContext2D.shadowOffsetX.get}            & Touch.force.get                                   \\* \midrule
\multicolumn{1}{|l|}{CanvasRenderingContext2D.roundRect}                    & \multicolumn{1}{l|}{OffscreenCanvasRenderingContext2D.shadowOffsetX.set}            & VisualViewport.height.get                         \\* \midrule
\multicolumn{1}{|l|}{CanvasRenderingContext2D.scale}                        & \multicolumn{1}{l|}{OffscreenCanvasRenderingContext2D.shadowOffsetY.get}            & VisualViewport.offsetLeft.get                     \\* \midrule
\multicolumn{1}{|l|}{CanvasRenderingContext2D.setTransform}                 & \multicolumn{1}{l|}{OffscreenCanvasRenderingContext2D.shadowOffsetY.set}            & VisualViewport.offsetTop.get                      \\* \midrule
\multicolumn{1}{|l|}{CanvasRenderingContext2D.shadowBlur.get}               & \multicolumn{1}{l|}{OffscreenCanvasRenderingContext2D.stroke}                       & VisualViewport.pageLeft.get                       \\* \midrule
\multicolumn{1}{|l|}{CanvasRenderingContext2D.shadowBlur.set}               & \multicolumn{1}{l|}{OffscreenCanvasRenderingContext2D.strokeRect}                   & VisualViewport.pageTop.get                        \\* \midrule
\multicolumn{1}{|l|}{CanvasRenderingContext2D.shadowColor.get}              & \multicolumn{1}{l|}{OffscreenCanvasRenderingContext2D.strokeStyle.get}              & VisualViewport.scale.get                          \\* \midrule
\multicolumn{1}{|l|}{CanvasRenderingContext2D.shadowColor.set}              & \multicolumn{1}{l|}{OffscreenCanvasRenderingContext2D.strokeStyle.set}              & VisualViewport.width.get                          \\* \midrule
\multicolumn{1}{|l|}{CanvasRenderingContext2D.shadowOffsetX.get}            & \multicolumn{1}{l|}{OffscreenCanvasRenderingContext2D.strokeText}                   & WebGL2RenderingContext.getExtension               \\* \midrule
\multicolumn{1}{|l|}{CanvasRenderingContext2D.shadowOffsetX.set}            & \multicolumn{1}{l|}{OffscreenCanvasRenderingContext2D.transform}                    & WebGL2RenderingContext.getInternalformatParameter \\* \midrule
\multicolumn{1}{|l|}{CanvasRenderingContext2D.shadowOffsetY.get}            & \multicolumn{1}{l|}{OffscreenCanvasRenderingContext2D.translate}                    & WebGL2RenderingContext.getParameter               \\* \midrule
\multicolumn{1}{|l|}{CanvasRenderingContext2D.shadowOffsetY.set}            & \multicolumn{1}{l|}{OffscreenCanvas.transferToImageBitmap}                          & WebGL2RenderingContext.getRenderbufferParameter   \\* \midrule
\multicolumn{1}{|l|}{CanvasRenderingContext2D.stroke}                       & \multicolumn{1}{l|}{PaintRenderingContext2D.arc}                                    & WebGL2RenderingContext.getShaderPrecisionFormat   \\* \midrule
\multicolumn{1}{|l|}{CanvasRenderingContext2D.strokeRect}                   & \multicolumn{1}{l|}{PaintRenderingContext2D.beginPath}                              & WebGL2RenderingContext.getSupportedExtensions     \\* \midrule
\multicolumn{1}{|l|}{CanvasRenderingContext2D.strokeStyle.get}              & \multicolumn{1}{l|}{PaintRenderingContext2D.clearRect}                              & WebGL2RenderingContext.makeXRCompatible           \\* \midrule
\multicolumn{1}{|l|}{CanvasRenderingContext2D.strokeStyle.set}              & \multicolumn{1}{l|}{PaintRenderingContext2D.closePath}                              & WebGLCompressedTextureASTC.getSupportedProfiles   \\* \midrule
\multicolumn{1}{|l|}{CanvasRenderingContext2D.strokeText}                   & \multicolumn{1}{l|}{PaintRenderingContext2D.drawMesh}                               & WebGLRenderingContext.getExtension                \\* \midrule
\multicolumn{1}{|l|}{CanvasRenderingContext2D.transform}                    & \multicolumn{1}{l|}{PaintRenderingContext2D.ellipse}                                & WebGLRenderingContext.getParameter                \\* \midrule
\multicolumn{1}{|l|}{CanvasRenderingContext2D.translate}                    & \multicolumn{1}{l|}{PaintRenderingContext2D.fill}                                   & WebGLRenderingContext.getRenderbufferParameter    \\* \midrule
\multicolumn{1}{|l|}{FeaturePolicy.features}                                & \multicolumn{1}{l|}{PaintRenderingContext2D.fillRect}                               & WebGLRenderingContext.getShaderPrecisionFormat    \\* \midrule
\multicolumn{1}{|l|}{Gamepad.id.get}                                        & \multicolumn{1}{l|}{PaintRenderingContext2D.fillStyle.get}                          & WebGLRenderingContext.getSupportedExtensions      \\* \midrule
\multicolumn{1}{|l|}{GPUAdapterInfo.architecture.get}                       & \multicolumn{1}{l|}{PaintRenderingContext2D.fillStyle.set}                          & WebGLRenderingContext.makeXRCompatible            \\* \midrule
\multicolumn{1}{|l|}{GPUAdapterInfo.description.get}                        & \multicolumn{1}{l|}{PaintRenderingContext2D.globalCompositeOperation.get}           & WheelEvent.deltaMode.get                          \\* \midrule
\multicolumn{1}{|l|}{GPUAdapterInfo.device.get}                             & \multicolumn{1}{l|}{PaintRenderingContext2D.globalCompositeOperation.set}           & WheelEvent.wheelDelta.get                         \\* \midrule
\multicolumn{1}{|l|}{GPUAdapterInfo.vendor.get}                             & \multicolumn{1}{l|}{PaintRenderingContext2D.isPointInPath}                          & WheelEvent.wheelDeltaX.get                        \\* \midrule
\multicolumn{1}{|l|}{History.length.get}                                    & \multicolumn{1}{l|}{PaintRenderingContext2D.isPointInStroke}                        & WheelEvent.wheelDeltaY.get                        \\* \midrule
\multicolumn{1}{|l|}{HTMLCanvasElement.captureStream}                       & \multicolumn{1}{l|}{PaintRenderingContext2D.lineTo}                                 & Window.devicePixelRatio.get                       \\* \midrule
\multicolumn{1}{|l|}{HTMLCanvasElement.getContext}                          & \multicolumn{1}{l|}{PaintRenderingContext2D.moveTo}                                 & Window.devicePixelRatio.set                       \\* \midrule
\multicolumn{1}{|l|}{HTMLCanvasElement.toBlob}                              & \multicolumn{1}{l|}{PaintRenderingContext2D.rect}                                   & Window.innerHeight.get                            \\* \midrule
\multicolumn{1}{|l|}{HTMLCanvasElement.toDataURL}                           & \multicolumn{1}{l|}{PaintRenderingContext2D.rotate}                                 & Window.innerHeight.set                            \\* \midrule
\multicolumn{1}{|l|}{HTMLMediaElement.canPlayType}                          & \multicolumn{1}{l|}{PaintRenderingContext2D.roundRect}                              & Window.innerWidth.get                             \\* \midrule
\multicolumn{1}{|l|}{HTMLVideoElement.webkitDecodedFrameCount.get}          & \multicolumn{1}{l|}{PaintRenderingContext2D.scale}                                  & Window.innerWidth.set                             \\* \midrule
\multicolumn{1}{|l|}{HTMLVideoElement.webkitDroppedFrameCount.get}          & \multicolumn{1}{l|}{PaintRenderingContext2D.shadowBlur.get}                         & Window.matchMedia                                 \\* \midrule
\multicolumn{1}{|l|}{InputDeviceCapabilities.firesTouchEvents.get}          & \multicolumn{1}{l|}{PaintRenderingContext2D.shadowBlur.set}                         & Window.orientation.get                            \\* \midrule
\multicolumn{1}{|l|}{Keyboard.getLayoutMap}                                 & \multicolumn{1}{l|}{PaintRenderingContext2D.shadowColor.get}                        & Window.outerHeight.get                            \\* \midrule
\multicolumn{1}{|l|}{MediaCapabilities.decodingInfo}                        & \multicolumn{1}{l|}{PaintRenderingContext2D.shadowColor.set}                        & Window.outerHeight.set                            \\* \midrule
\multicolumn{1}{|l|}{MediaCapabilities.encodingInfo}                        & \multicolumn{1}{l|}{PaintRenderingContext2D.shadowOffsetX.get}                      & Window.outerWidth.get                             \\* \midrule
\multicolumn{1}{|l|}{MediaDevices.enumerateDevices}                         & \multicolumn{1}{l|}{PaintRenderingContext2D.shadowOffsetX.set}                      & Window.outerWidth.set                             \\* \midrule
\multicolumn{1}{|l|}{MediaDevices.getUserMedia}                             & \multicolumn{1}{l|}{PaintRenderingContext2D.shadowOffsetY.get}                      & Window.pageXOffset.get                            \\* \midrule
\multicolumn{1}{|l|}{MediaRecorder.audioBitsPerSecond.get}                  & \multicolumn{1}{l|}{PaintRenderingContext2D.shadowOffsetY.set}                      & Window.pageXOffset.set                            \\* \midrule
\multicolumn{1}{|l|}{MediaRecorder.mimeType.get}                            & \multicolumn{1}{l|}{PaintRenderingContext2D.stroke}                                 & Window.pageYOffset.get                            \\* \midrule
\multicolumn{1}{|l|}{MediaRecorder.videoBitsPerSecond.get}                  & \multicolumn{1}{l|}{PaintRenderingContext2D.strokeRect}                             & Window.pageYOffset.set                            \\* \midrule
\multicolumn{1}{|l|}{MouseEvent.screenX.get}                                & \multicolumn{1}{l|}{PaintRenderingContext2D.strokeStyle.get}                        & Window.screenLeft.get                             \\* \midrule
\multicolumn{1}{|l|}{MouseEvent.screenY.get}                                & \multicolumn{1}{l|}{PaintRenderingContext2D.strokeStyle.set}                        & Window.screenLeft.set                             \\* \midrule
\multicolumn{1}{|l|}{Navigator.appVersion.get}                              & \multicolumn{1}{l|}{PaintRenderingContext2D.transform}                              & Window.screenTop.get                              \\* \midrule
\multicolumn{1}{|l|}{Navigator.cookieEnabled.get}                           & \multicolumn{1}{l|}{PaintRenderingContext2D.translate}                              & Window.screenTop.set                              \\* \midrule
\multicolumn{1}{|l|}{Navigator.deviceMemory.get}                            & \multicolumn{1}{l|}{PaintWorkletGlobalScope.devicePixelRatio.get}                   & Window.screenX.get                                \\* \midrule
\multicolumn{1}{|l|}{Navigator.doNotTrack.get}                              & \multicolumn{1}{l|}{Path2D.arc}                                                     & Window.screenX.set                                \\* \midrule
\multicolumn{1}{|l|}{Navigator.getUserMedia}                                & \multicolumn{1}{l|}{Path2D.closePath}                                               & Window.screenY.get                                \\* \midrule
\multicolumn{1}{|l|}{Navigator.hardwareConcurrency.get}                     & \multicolumn{1}{l|}{Path2D.ellipse}                                                 & Window.screenY.set                                \\* \midrule
\multicolumn{1}{|l|}{Navigator.javaEnabled}                                 & \multicolumn{1}{l|}{Path2D.lineTo}                                                  & Window.scrollX.get                                \\* \midrule
\multicolumn{1}{|l|}{Navigator.language.get}                                & \multicolumn{1}{l|}{Path2D.moveTo}                                                  & Window.scrollX.set                                \\* \midrule
\multicolumn{1}{|l|}{Navigator.languages.get}                               & \multicolumn{1}{l|}{Path2D.rect}                                                    & Window.scrollY.get                                \\* \midrule
\multicolumn{1}{|l|}{Navigator.maxTouchPoints.get}                          & \multicolumn{1}{l|}{Path2D.roundRect}                                               & Window.scrollY.set                                \\* \midrule
\multicolumn{1}{|l|}{Navigator.mimeTypes.get}                               & \multicolumn{1}{l|}{PaymentRequest.canMakePayment}                                  & WorkerNavigator.appVersion.get                    \\* \midrule
\multicolumn{1}{|l|}{Navigator.pdfViewerEnabled.get}                        & \multicolumn{1}{l|}{PaymentRequest.hasEnrolledInstrument}                           & WorkerNavigator.deviceMemory.get                  \\* \midrule
\multicolumn{1}{|l|}{Navigator.platform.get}                                & \multicolumn{1}{l|}{Plugin.description.get}                                         & WorkerNavigator.hardwareConcurrency.get           \\* \midrule
\multicolumn{1}{|l|}{Navigator.plugins.get}                                 & \multicolumn{1}{l|}{Plugin.filename.get}                                            & WorkerNavigator.language.get                      \\* \midrule
\multicolumn{1}{|l|}{Navigator.productSub.get}                              & \multicolumn{1}{l|}{Plugin.name.get}                                                & WorkerNavigator.languages.get                     \\* \midrule
\multicolumn{1}{|l|}{NavigatorUAData.brands.get}                            & \multicolumn{1}{l|}{PushManager.supportedContentEncodings.get}                      & WorkerNavigator.platform.get                      \\* \midrule
\multicolumn{1}{|l|}{NavigatorUAData.getHighEntropyValues}                  & \multicolumn{1}{l|}{RTCIceCandidate.address.get}                                    & WorkerNavigator.userAgent.get                     \\* \midrule
\multicolumn{1}{|l|}{NavigatorUAData.mobile.get}                            & \multicolumn{1}{l|}{}                                                               &                                                   \\* \bottomrule
\end{longtable}
\setlength{\tabcolsep}{6pt} % Restore to default
\end{footnotesize}

\section{List of bytecode tokens}\label{sec:appendix2}
This subsection presents a list of JavaScript bytecode tokens extracted from all scripts collected across a dataset of 100,000 websites. These tokens represent the low-level instructions generated by the V8 JavaScript engine (used in Chromium-based browsers) after parsing and compiling JavaScript code.

Each token corresponds to a specific operation (e.g., loading a variable, calling a function, reading a property) in the V8 interpreter’s bytecode. By analyzing the frequency and patterns of these tokens, we can gain insights into script behavior and potentially identify patterns related to tracking, fingerprinting, or other complex client-side logic.

For reference, a complete list of available bytecode instructions (tokens) can also be found in the V8 source code, specifically in the file bytecode.h.

\begin{footnotesize}
\begin{longtable}[c]{@{}llll@{}}
\caption{Bytecode tokens}\label{tab:tokens} \\
\toprule
\multicolumn{4}{c}{\textbf{Bytecode tokens}} \\
\midrule
\endfirsthead

\toprule
\multicolumn{4}{c}{\textbf{Bytecode tokens (continued)}} \\
\midrule
\endhead

\bottomrule
\endfoot

\bottomrule
\endlastfoot
CreateFunctionContext & SetKeyedProperty.Wide & LdaLookupGlobalSlot & PushContext \\
GetKeyedProperty.Wide & CreateFunctionContext.Wide & Ldar & GetNamedProperty.Wide \\
JumpIfTrue.Wide & StaCurrentContextSlot & CallUndefinedReceiver.Wide & JumpIfUndefined.Wide \\
CreateMappedArguments & ShiftRightSmi.Wide & ForInPrepare.Wide & Star1 \\
TestGreaterThanOrEqual.Wide & ForInNext.Wide & LdaCurrentContextSlot & Add.Wide \\
JumpIfNotNullConstant & JumpIfToBooleanFalse & CallProperty1.Wide & Div.Wide \\
Star3 & BitwiseAnd.Wide & LdaLookupContextSlot & GetNamedProperty \\
TestLessThan.Wide & ModSmi.ExtraWide & JumpIfToBooleanTrue & CallProperty0.Wide \\
Star.Wide & LdaImmutableContextSlot & TestEqualStrict.Wide & Ldar.Wide \\
Star4 & ShiftLeft.Wide & Mov.Wide & CallProperty2 \\
AddSmi.Wide & JumpIfNotUndefined.Wide & Return & BitwiseOr.Wide \\
BitwiseXorSmi & CreateEmptyObjectLiteral & Dec.Wide & BitwiseXorSmi.Wide \\
Star5 & CallProperty2.Wide & StaLookupSlot & Star6 \\
Inc.Wide & BitwiseXorSmi.ExtraWide & CallUndefinedReceiver1 & Sub.Wide \\
ConstructWithSpread & CallProperty1 & TestLessThanOrEqual.Wide & BitwiseXor.Wide \\
LdaImmutableCurrentContextSlot & MulSmi.Wide & GetNamedPropertyFromSuper & LdaConstant \\
TestGreaterThan.Wide & StaGlobal.Wide & Star0 & ShiftRight.Wide \\
CloneObject.Wide & LdaZero & LdaGlobal.Wide & CreateCatchContext.Wide \\
GetKeyedProperty & Construct.Wide & PushContext.Wide & SetNamedProperty \\
Mul.Wide & CallRuntime.Wide & CreateArrayLiteral & CallProperty.Wide \\
CallWithSpread.Wide & CreateClosure & Mod & JumpIfJSReceiver.Wide \\
LdaSmi & ShiftRight & Mod.Wide & TestEqualStrict \\
CallUndefinedReceiver0 & JumpIfUndefinedOrNull.Wide & JumpIfFalse & DefineKeyedOwnProperty \\
LdaLookupSlotInsideTypeof & SetKeyedProperty & ToBoolean & GetTemplateObject.Wide \\
Add & CreateCatchContext & JumpIfNotUndefinedConstant & Jump \\
LdaTheHole & ShiftLeftSmi.Wide & DivSmi & SetPendingMessage \\
ShiftRightLogical.Wide & CallUndefinedReceiver & PopContext & CallJSRuntime \\
TestTypeOf & DeletePropertyStrict & ThrowReferenceErrorIfHole.Wide & JumpIfTrue \\
JumpIfToBooleanFalseConstant & CallUndefinedReceiver1.ExtraWide & LdaUndefined & SwitchOnSmiNoFeedback \\
Add.ExtraWide & LdaNull & ReThrow & CallUndefinedReceiver2.ExtraWide \\
Star7 & TestReferenceEqual & SetKeyedProperty.ExtraWide & Star8 \\
StaContextSlot & ShiftRightLogicalSmi.Wide & StaInArrayLiteral & JumpIfNotNull \\
BitwiseNot.Wide & Star2 & JumpIfNull & JumpLoop.ExtraWide \\
Star & TypeOf & JumpIfTrueConstant.Wide & CallProperty0 \\
ThrowReferenceErrorIfHole & JumpConstant.Wide & Mov & FindNonDefaultConstructorOrConstruct \\
PopContext.Wide & CreateObjectLiteral & ThrowIfNotSuperConstructor & ToObject.Wide \\
DefineNamedOwnProperty & ConstructForwardAllArgs & ForInEnumerate.Wide & Star10 \\
CreateBlockContext & ForInContinue.Wide & Star9 & LdaImmutableContextSlot.Wide \\
ForInStep.Wide & Star11 & LdaImmutableCurrentContextSlot.Wide & LdaModuleVariable.Wide \\
CallProperty & Div & LdaModuleVariable & Construct \\
Negate & StaModuleVariable & DeletePropertySloppy & StaContextSlot.Wide \\
JumpIfNull.Wide & TestGreaterThan & LdaContextSlot.Wide & LdaLookupGlobalSlot.Wide \\
JumpIfUndefinedOrNull & ToString & LdaLookupGlobalSlotInsideTypeof & ToObject \\
CreateRestParameter & Exp & ForInEnumerate & GetIterator \\
LdaLookupGlobalSlotInsideTypeof.Wide & ForInPrepare & JumpIfJSReceiver & Debugger \\
ForInContinue & CallRuntime & ExpSmi & CreateArrayFromIterable \\
CreateWithContext & JumpIfUndefined & JumpIfNullConstant & TestReferenceEqual.Wide \\
Star12 & JumpIfUndefinedConstant & DeletePropertyStrict.Wide & Star13 \\
LdaCurrentContextSlot.Wide & CallAnyReceiver.Wide & TestUndefined & StaCurrentContextSlot.Wide \\
ShiftRightSmi.ExtraWide & LogicalNot & ThrowSuperAlreadyCalledIfNotHole & ShiftLeftSmi.ExtraWide \\
Star15 & ThrowSuperNotCalledIfHole & StaModuleVariable.Wide & Star14 \\
BitwiseOrSmi.Wide & InvokeIntrinsic.Wide & ForInStep & TestNull \\
SuspendGenerator.Wide & JumpLoop & SwitchOnGeneratorState & ResumeGenerator.Wide \\
Inc & InvokeIntrinsic & SwitchOnSmiNoFeedback.Wide & LdaContextSlot \\
SuspendGenerator & Exp.Wide & CreateEmptyArrayLiteral & ResumeGenerator \\
LdaLookupSlot.Wide & LdaFalse & SetNamedProperty.Wide & JumpIfNotNull.Wide \\
LdaTrue & CallUndefinedReceiver2.Wide & LdaLookupContextSlot.Wide & CallUndefinedReceiver2 \\
CreateRegExpLiteral & LdaLookupContextSlotInsideTypeof & TestUndetectable & CloneObject \\
GetNamedPropertyFromSuper.Wide & LdaGlobal & CreateObjectLiteral.Wide & StaLookupSlot.Wide \\
Sub & CallUndefinedReceiver0.Wide & DefineNamedOwnProperty.ExtraWide & ShiftLeft \\
DivSmi.Wide & SetNamedProperty.ExtraWide & TestLessThan & ToName \\
GetNamedProperty.ExtraWide & TestLessThanOrEqual & DefineKeyedOwnPropertyInLiteral & CreateObjectLiteral.ExtraWide \\
ShiftLeftSmi & GetTemplateObject & CallUndefinedReceiver0.ExtraWide & SubSmi \\
CreateClosure.Wide & LdaGlobal.ExtraWide & BitwiseAnd & CreateBlockContext.Wide \\
CallProperty1.ExtraWide & ShiftRightSmi & LdaConstant.Wide & Construct.ExtraWide \\
LdaGlobalInsideTypeof & CallUndefinedReceiver1.Wide & CallProperty2.ExtraWide & TestInstanceOf \\
Negate.Wide & CreateEvalContext.Wide & MulSmi & StaGlobal \\
JumpIfFalseConstant.Wide & ShiftRightLogicalSmi & CreateArrayLiteral.Wide & DeletePropertySloppy.Wide \\
SwitchOnSmiNoFeedback.ExtraWide & ModSmi & StaInArrayLiteral.Wide & Throw \\
DefineKeyedOwnProperty.Wide & GetKeyedProperty.ExtraWide & AddSmi & JumpIfUndefinedOrNullConstant \\
DefineKeyedOwnProperty.ExtraWide & BitwiseOr & BitwiseXor & StaInArrayLiteral.ExtraWide \\
BitwiseAndSmi.ExtraWide & GetIterator.Wide & CreateArrayLiteral.ExtraWide & BitwiseAndSmi.Wide \\
DefineNamedOwnProperty.Wide & CreateEmptyArrayLiteral.ExtraWide & ToNumeric & DefineKeyedOwnPropertyInLiteral.Wide \\
ConstructWithSpread.Wide & Dec & AddSmi.ExtraWide & BitwiseAndSmi \\
CreateEmptyArrayLiteral.Wide & CallUndefinedReceiver.ExtraWide & CallProperty.ExtraWide & LdaSmi.ExtraWide \\
TestEqual.Wide & CallJSRuntime.Wide & ShiftRightLogical & SubSmi.ExtraWide \\
CallProperty0.ExtraWide & LdaSmi.Wide & LdaLookupSlot & GetSuperConstructor \\
TestGreaterThanOrEqual & CallRuntimeForPair & CreateClosure.ExtraWide & SubSmi.Wide \\
CallAnyReceiver & LdaConstant.ExtraWide & CreateUnmappedArguments & ToNumeric.Wide \\
FindNonDefaultConstructorOrConstruct.Wide & TestIn & CreateRegExpLiteral.Wide & ThrowIfNotSuperConstructor.Wide \\
ToBooleanLogicalNot & TestInstanceOf.Wide & JumpIfToBooleanFalseConstant.Wide & Mul \\
ModSmi.Wide & JumpIfToBooleanTrueConstant.Wide & ToNumber & CallWithSpread \\
CreateWithContext.Wide & JumpIfNotUndefined & JumpIfToBooleanTrue.Wide & CallRuntimeForPair.Wide \\
BitwiseOrSmi & JumpIfFalse.Wide & TestEqual.ExtraWide & JumpLoop.Wide \\
Jump.Wide & BitwiseAnd.ExtraWide & MulSmi.ExtraWide & DivSmi.ExtraWide \\
TestEqualStrict.ExtraWide & TestEqual & ToNumber.Wide & CreateRegExpLiteral.ExtraWide \\
BitwiseNot & BitwiseOrSmi.ExtraWide & GetTemplateObject.ExtraWide & JumpIfTrueConstant \\
CreateEvalContext & CloneObject.ExtraWide & JumpIfToBooleanTrueConstant & JumpConstant \\
LdaGlobalInsideTypeof.Wide & JumpIfFalseConstant & JumpIfToBooleanFalse.Wide & \\* \bottomrule
\end{longtable}
\end{footnotesize}

\end{document}